\documentclass[preprint]{elsarticle}

\usepackage{graphicx}
\usepackage{amsmath,amssymb}
\usepackage[body={6in,8.5in}]{geometry}

\usepackage{subfigure}
\usepackage{color}
\usepackage{url}
\usepackage{hyperref}

\def\cL{\mathcal{L}}
\def\cN{\mathcal{N}}

\def\HEXplanhexa{http://www.youtube.com/watch?v=5OkFRs1Sczw}
\def\HEXplanhexb{http://www.youtube.com/watch?v=gfsXKo1AQ0g}
\def\HEXplanhexc{http://www.youtube.com/watch?v=6A0_ZDcOc_8}

\def\HEXlocringa{http://www.youtube.com/watch?v=Z_MGpmm0uHk}
\def\HEXcirchexs{https://www.youtube.com/watch?v=KM8cO2Tc4nw}
\def\HEXcirchexa{http://www.youtube.com/watch?v=-9H_6-lQHAY}

\def\HEXsubfullym{https://www.youtube.com/watch?v=jpezS-GoMjA}
\def\HEXsubfullyp{https://www.youtube.com/watch?v=6mqPUwYeXiA}
\def\HEXsubplanarm{https://www.youtube.com/watch?v=aVp0XupDNXs}
\def\HEXsubplanarp{https://www.youtube.com/watch?v=MbewDIBwXQo}

\numberwithin{equation}{section}

\begin{document}

\title{Two-dimensional localized structures in harmonically forced oscillatory systems}

\author[a]{Y.-P. Ma\corref{cor1}}
\ead{yiping.m@gmail.com}

\author[b]{E. Knobloch}
\ead{knobloch@berkeley.edu}

\cortext[cor1]{Corresponding author}
\address[a]{Department of Applied Mathematics, University of Colorado, Boulder, CO 80309, USA}
\address[b]{Department of Physics, University of California, Berkeley, CA 94720, USA}

\begin{abstract}
Two-dimensional spatially localized structures in the complex Ginzburg-Landau equation with 1:1 resonance are studied near the simultaneous occurrence of a steady front between two spatially homogeneous equilibria and a supercritical Turing bifurcation on one of them. The bifurcation structures of steady circular fronts and localized target patterns are computed in the Turing-stable and Turing-unstable regimes. In particular, localized target patterns grow along the solution branch via ring insertion at the core in a process reminiscent of defect-mediated snaking in one spatial dimension. Axisymmetric oscillons on these solution branches are found to be stable over a wide parameter interval, and subject to various types of instability otherwise. Direct numerical simulations reveal novel depinning dynamics of localized target patterns in the radial direction, and of circular and planar localized hexagonal patterns in the fully two-dimensional system.
\end{abstract}


\date{\today}

\maketitle

\section{Introduction}\label{sec:intro}

For several decades, two-dimensional (2D) localized structures have been a frequent subject of experimental, numerical and analytical studies~\cite{PuBoAm10}. A particularly interesting class of localized structures, known as localized states (LS), results from the coexistence of two stationary states, either of which can fill the entire spatial domain~\cite{K:08}. These states may take the form of spatially homogeneous equilibria or spatially periodic patterns. A stable pattern can be created from an equilibrium $A_0$ by the Turing instability, which can be either supercritical or subcritical. In the supercritical case, since the equilibrium $A_0$ is unstable while the pattern is stable, the latter invades the former via a pulled front and thus no steady LS are expected~\cite{vS03}; Ref.~\cite{CsMi99} describes the invasion of a homogeneous state by a hexagonal pattern of this type. In the subcritical case, $A_0$ and the pattern can both be stable and the front between them (known as a pushed front) may remain pinned to the pattern over a range of parameter values \cite{Pom:86,JPMDB94}. This pinning mechanism underlies much recent work on both 1D and 2D LS, notably in the quadratic-cubic Swift-Hohenberg equation (SH23). The bifurcation structures and temporal dynamics of 1D LS in SH23 are now well understood~\cite{BuKn:06}: stationary reflection-symmetric LS are located on a pair of intertwined branches, one of which corresponds to states with maxima in the center and the other to states with minima in the center. As one proceeds along either branch these states grow in length via the repeated nucleation of a pair of wavelengths, one on either side of the LS, in a process reflected in the back-and-forth oscillations of the branches and referred to as standard homoclinic snaking. The two LS branches are in addition interconnected by an infinite number of cross-links (known as ``ladders'') formed of stationary but asymmetric LS. These states, together with the associated multipulse states, coexist within a parameter interval referred to as the snaking or pinning region. Outside this interval the LS grow or decay in time via a process referred to as depinning, i.e., through the nucleation or destruction of cells at the location of the fronts in a time-dependent manner, while leaving the interior cells undisturbed. See \cite{Knobloch15} for a recent review.

More recently, 2D LS in SH23 have also been investigated. In Ref.~\cite{McCS10} localized $d$-dimensional axisymmetric solutions, hereafter referred to as localized target patterns, are studied using numerical continuation. The time-independent axisymmetric SH23 in $d$ dimensions is treated as a spatial dynamical system in the radial coordinate $r$, with $d$ treated as an additional continuous parameter. For localized target patterns with a large radius, the radial profile of the front between the target and the equilibrium $A_0$ resembles a 1D pushed front. However, in contrast to 1D, near $r=0$ (known as the core region) the target typically has a larger amplitude than at large $r$ (known as the far field). As a result new rings can be nucleated both in the core and at the front, implying that the snaking behavior of localized target patterns for $d>1$ will differ from standard homoclinic snaking at $d=1$. In particular, when the radius is sufficiently large, new rings are nucleated exclusively in the core. A different type of localized structure, stationary spatially localized hexagonal patterns, are studied in Ref.~\cite{LSAC08} using numerical continuation. The existence of these solutions follows from a spatial conservation law for the time-independent SH23 in 2D that generalizes a conserved spatial Hamiltonian present in 1D. Two classes of such states are known, hereafter referred to as planar and fully localized hexagonal patterns. The former are localized in 1D, forming a hexagon-filled stripe connected to the background state by a pair of planar fronts, while the latter are bounded by a hexagonal front and localized in 2D. Both structures grow via the nucleation of additional cells along the bounding fronts, in a process similar to standard homoclinic snaking. A novel feature of the planar case is the presence of secondary snaking branches, corresponding to the nucleation of individual cells {\it along} either front, between successive oscillations of the primary snaking branch that reflect the addition of complete rows of cells along the fronts. This growth mechanism is also observed along the snaking curve of fully localized hexagons. Outside the snaking region time-dependent calculations show that these almost planar solutions evolve by losing or gaining cells one-by-one in a manner that recapitulates their growth along the snaking branch of time-independent solutions.

\begin{figure}
\center
\begin{tabular}{cc}
\includegraphics[width=0.48\textwidth]{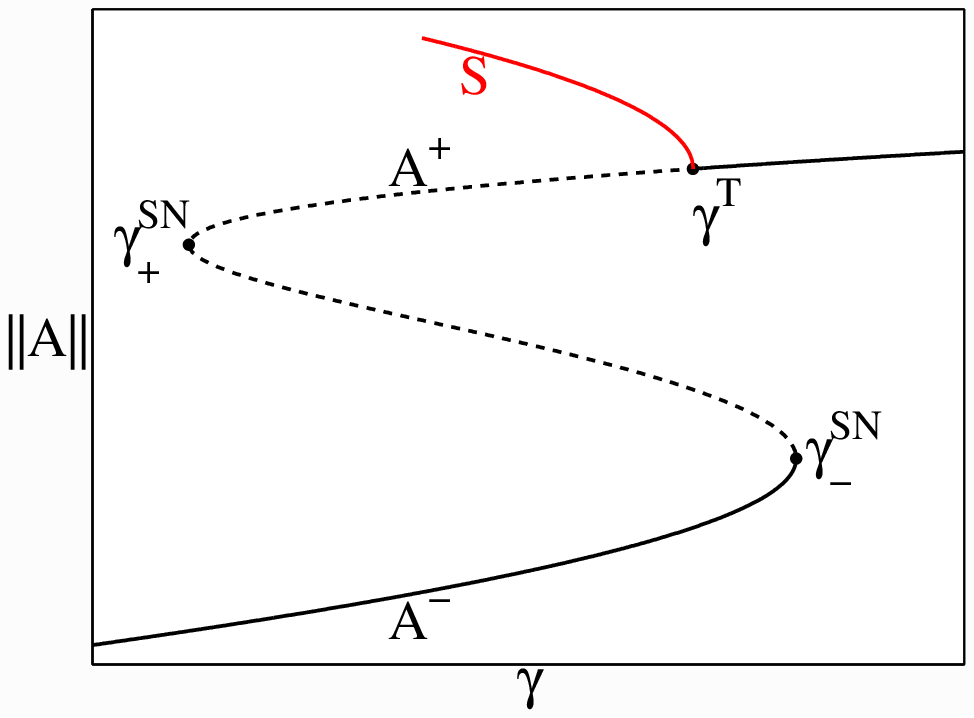} &
\includegraphics[width=0.48\textwidth]{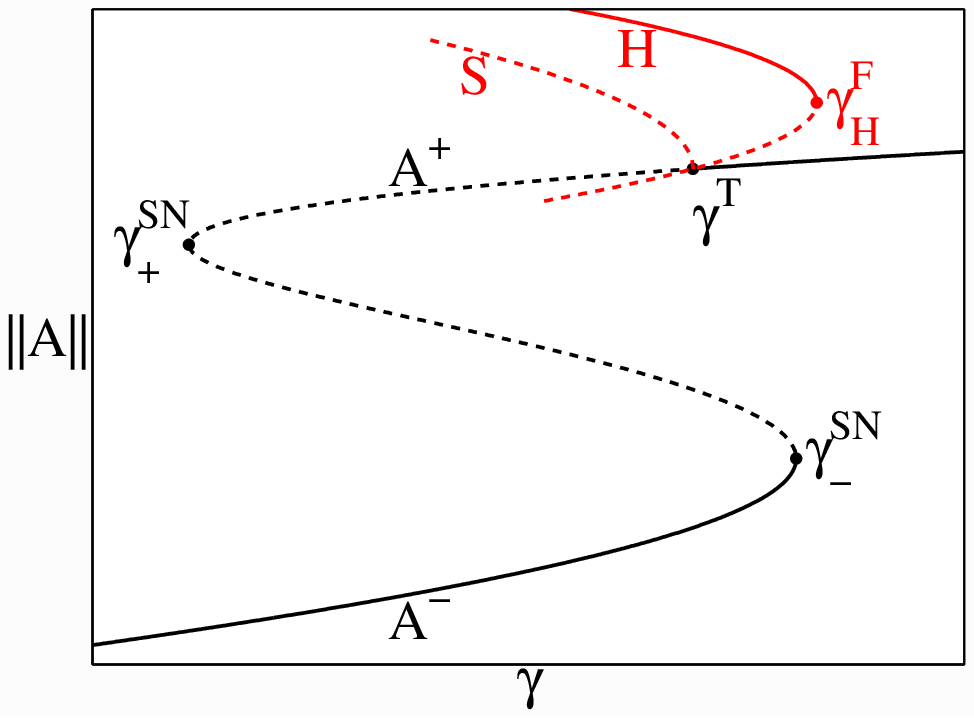} \\
(a) & (b)
\end{tabular}
\caption{Sketch of the bifurcation diagrams of the building blocks of (a) 1D and (b) 2D LS in the 1:1 FCGLE (\ref{eq:FCGL1to1}). Two equilibria, labeled $A^\pm$ and separated by a pair of saddle-node bifurcations $\gamma_\pm^{SN}$, may be simultaneously stable. The Turing bifurcation on $A^+$ at $\gamma^T$ produces a branch of stripes $S$ in 1D, and branches of stripes $S$ and hexagons $H$ in 2D. The solid (dashed) segments are stable (unstable).} \label{fig:HEX_elements}
\end{figure}

In the above example one of the two states comprising the LS is homogeneous in space while the other is a steady, spatially periodic Turing pattern. In other PDEs the two competing states may be both homogeneous states or both stationary patterns. A notable pattern-forming PDE exhibiting bistability between homogeneous states is the forced complex Ginzburg-Landau equation (FCGLE) with 2:1 resonance. This is an equation for a complex scalar amplitude $A=U+iV$~\cite{CE:92} which describes the effect of subharmonic forcing on a Hopf bifurcation to a homogeneous oscillation in spatially extended systems~\cite{ANR14}. This PDE possesses an up-down symmetry $A\rightarrow-A$ and exhibits a parameter regime where the two ``equivalent'' equilibria $\pm A_0$ related by this symmetry are simultaneously stable. In 1D a codimension-zero family of steady Ising fronts between $\pm A_0$ is generically present, and may undergo a pitchfork bifurcation into traveling Bloch fronts~\cite{BYK:08}; near this Ising-Bloch transition the domain walls can exhibit rich spatiotemporal dynamics~\cite{GoCoWa15}. In Ref.~\cite{GCMO07}, a dynamical theory for the domain wall between $\pm A_0$ in 2D leads to a characterization of different growth regimes and the prediction of the existence of stable ``droplets''. In this paper we study instead the 1:1 FCGLE
\begin{equation}\label{eq:FCGL1to1}
\frac{\partial A}{\partial t}=(1+i\alpha)\nabla^2A+(\mu+i\nu)A-(1+i\beta)|A|^2A+\gamma,
\end{equation}
which exhibits, under appropriate conditions, bistability between two equilibria $A^+=U^++iV^+$ and $A^-=U^-+iV^-$ belonging to the upper and lower branches of an S-shaped bifurcation diagram (Fig.~\ref{fig:HEX_elements}). In 1D this regime generically exhibits a codimension-one family of steady Bloch fronts between $A^\pm$ when both $A^\pm$ are stable. Steady LS known as type-I LS can be constructed by assembling two such fronts back-to-back. Recent work has focused on new dynamical phenomena that arise when $A^+$ becomes Turing-unstable, including the snaking behavior and depinning dynamics of steady localized Turing patterns known as type-II LS~\cite{MaThesis}. In contrast to standard homoclinic snaking, type-II LS grow along a single snaking branch by nucleating new cells at the center of the wavetrain in a process termed defect-mediated snaking (DMS)~\cite{MBK10}. Outside the snaking region, type-II LS depin by creating or destroying cells in the interior of the wavetrain via successive phase slips, as studied quantitatively in Ref.~\cite{MaKn12}. In the present paper, we characterize 2D LS in the $(x,y)$ plane in the parameter regime where type-I and type-II LS are found in 1D.

The bifurcation diagrams of the building blocks of these 1D and 2D LS are shown schematically in Fig.~\ref{fig:HEX_elements} in terms of a suitable norm $\|A\|$ as a function of the control parameter $\gamma$. In 1D, a branch of stripes $S$ bifurcates supercritically from $A^+$ at the Turing bifurcation $\gamma^T$, and may become bistable with $A^-$ to form type-II LS (Fig.~\ref{fig:HEX_elements}(a)). In 2D, the Turing bifurcation generates, in addition to $S$, a pair of hexagonal patterns, hereafter referred to as $H^{\pm}$, that differ in whether the hexagons superposed on $A^+$ are composed of bumps or holes \cite{Golubitsky1984}. As a result the stripe pattern is unstable to hexagonal perturbations even when it bifurcates supercritically. The $H^{\pm}$ bifurcate transcritically and are likewise both unstable. However, in general one or the other of the $H^{\pm}$ folds back at $\gamma_H^F>\gamma^T$ and acquires stability as indicated in Fig.~\ref{fig:HEX_elements}(b). This is the case in the present case as well.

We begin by focusing on one-pulse solutions in which the region occupied by the $A^+$ state forms a single connected component embedded in the $A^-$ state, although we briefly comment on ``reciprocal'' localized states in which the $A^-$ state is embedded in the $A^+$ state. When the connected component is finite, the interface separating $A^\pm$ forms a closed curve, and we refer to the region inside (outside) this curve as the inner (outer) region. For most of the localized states considered in this paper, this curve takes the shape of a perfect circle, the only shape that respects the rotational symmetry of Eq.~(\ref{eq:FCGL1to1}). This behavior is reminiscent of variational systems that tend to minimize a surface energy, even though Eq.~(\ref{eq:FCGL1to1}) is nonvariational. In \S\ref{sec:circtw} we study the simplest case, the case of circular fronts between $A^\pm$. Next, we consider cases where the $A^+$ state is Turing-unstable, leading to a patterned state near $A^+$. As shown in \S\ref{sec:locring} the steady-state solutions that result may take the form of an (approximately) periodic structure in the radial direction, and we find localized target patterns of this type using numerical continuation. In \S\ref{sec:circhex} we address the more complicated case where the pattern has hexagonal structure, and use direct numerical simulation (DNS) to identify a new type of LS consisting of localized hexagons that are surrounded by a ring, with this compound structure embedded in $A^-$. We refer to these states as {\it circular localized hexagonal patterns}.

In order to study axisymmetric solutions, we write the 1:1 FCGLE (\ref{eq:FCGL1to1}) in polar coordinates $(r,\theta)$,
\begin{equation}\label{eq:FCGL_polar}
A_t=(1+i\alpha)\left(\partial_{rr}+\frac{1}{r}\partial_r+\frac{1}{r^2}\partial_{\theta\theta}\right)A+f(A),
\end{equation}
where $f(A)\equiv(\mu+i\nu)A-(1+i\beta)|A|^2A+\gamma$. Steady axisymmetric solutions satisfy Eq.~(\ref{eq:FCGL_polar}) with $\partial_tA=\partial_\theta A=0$, i.e.,
\begin{equation}\label{eq:circfr_steady2}
\left(\partial_{rr}+\frac{1}{r}\partial_r\right)A+\frac{1}{1+i\alpha}f(A)=0.
\end{equation}
Equation~(\ref{eq:circfr_steady2}) can be readily generalized from 2D to $d$ dimensions,
\begin{equation}\label{eq:circfr_steadyn}
\left(\partial_{rr}+\frac{d-1}{r}\partial_r\right)A+\frac{1}{1+i\alpha}f(A)=0,
\end{equation}
where $d$ is treated as a continuous parameter \cite{McCS10}. The boundary conditions for axisymmetric solutions are taken to be Neumann boundary conditions (NBC), i.e., $A_r=0$ at $r=0,\infty$, independently of $d$, or in terms of the real and imaginary parts of $A$,
\begin{equation}\label{eq:NBC}
U_r=V_r=0,\quad r=0,\infty.
\end{equation}

The $L^2$-norm of a $d$-dimensional axisymmetric solution $A(r)$ on a circular domain $r\in[0,R]$ is measured using either the 1D radial norm
\begin{equation}\label{eq:norm-1d}
\|A\|_x:=\sqrt{\frac{1}{R}\int_0^R|A(r)|^2dr},
\end{equation}
or the $d$-dimensional radial norm
\begin{equation}\label{eq:norm-dsym}
\|A\|_{r,d}:=\sqrt{\int_0^R|A(r)|^2r^{d-1}dr\left/\int_0^Rr^{d-1}dr\right.}.
\end{equation}
Although the $d$-dimensional radial norm is the physical $L^2$-norm for a multi-dimensional solution, the 1D radial norm weights all parts of the solution profile equally and hence better represents the profile changes near $r=0$. The $L^2$-norm for a 2D solution $A(x,y)$ (not necessarily axisymmetric) on a 2D domain $\Omega$ is defined as
\begin{equation}\label{eq:norm-2gen}
\|A\|_{(x,y)}:=\sqrt{\iint_\Omega|A(x,y)|^2dxdy\left/\iint_\Omega dxdy\right.}.
\end{equation}
This fully 2D norm is compatible with the $d$-dimensional radial norm because Eq.~(\ref{eq:norm-dsym}) with $d=2$ is identical to Eq.~(\ref{eq:norm-2gen}) when $A$ depends on $r$ only and $\Omega$ is a circular domain of radius $R$. For a localized state consisting of an inner pattern $A^i$ of radius $\rho$ embedded in an outer pattern $A^o$ on a domain of radius $R$ ($R>\rho$), any of the above $L^2$-norms generates a monotonic function of $\rho$ that is well approximated by
\begin{equation}\label{eq:norm-rho}
\|A\|^2=\|A^o\|^2+\frac{\rho^d}{R^d}\left(\|A^i\|^2-\|A^o\|^2\right).
\end{equation}

The spectral stability of an axisymmetric LS $A_0(r)$ depends on its stability with respect to perturbations proportional to $e^{im\theta}$, or equivalently with azimuthal wavenumber $m$, where $m\in\mathbb{Z}$. Hereafter we refer to this perturbation via its symmetry group as a $\mathbb{D}_m$ perturbation, where $\mathbb{D}_m$ denotes the dihedral group of order $m$. Thus $\mathbb{D}_0$ represents continuous rotations, $\mathbb{D}_1$ represents the trivial rotation by $2\pi$, while $\mathbb{D}_m$ for $m\geq2$ is generated by rotations by $2\pi/m$ together with the reflections $\theta\to -\theta$. The $\mathbb{D}_m$ stability of $A_0(r)$ is given by the spectrum of the linearized FCGL operator with $\partial_{\theta\theta}=-m^2$, namely
\begin{equation*}
\cL=\left[
\begin{array}{cc}
 \mu  & -\nu  \\
 \nu  & \mu
\end{array}
\right]+\left[
\begin{array}{cc}
 1 & -\alpha  \\
 \alpha  & 1
\end{array}
\right]\left(\partial _{rr}+\frac{1}{r}\partial_r-\frac{m^2}{r^2}\right)-\left[
\begin{array}{cc}
 1 & -\beta  \\
 \beta  & 1
\end{array}
\right]\left[
\begin{array}{cc}
 3 U_0^2+V_0^2 & 2 U_0 V_0 \\
 2 U_0 V_0 & U_0^2+3 V_0^2
\end{array}
\right],
\end{equation*}
with the boundary conditions given by Eq.~(\ref{eq:NBC}).

In addition to the above fully localized states in which either the $A^+$ state or the patterned state occupies a finite region, there exist planar localized states in which these states occupy a region that is infinite along a certain direction. If such localized states are computed on a rectangle with periodic boundary conditions (PBC), we may suppose that the solutions are localized in, say, the $x$-direction, and periodic in the orthogonal direction, namely the $y$-direction. Two solutions of this type, planar fronts between $A^+$ and $A^-$ and localized stripe patterns can be constructed by extending Type-I LS and Type-II LS in the $y$-direction. Because these states are invariant under translation in $y$, both satisfy the same PDE as their 1D counterparts and so will not be discussed further. Instead we focus in \S\ref{sec:planhex} on nontrivial examples in which the inner pattern is itself periodic in 2D. We conclude in \S\ref{sec:end-HEX} with some open questions.

As in Refs.~\cite{MBK10,MaKn12} we fix three of the parameters in Eq.~\ref{eq:FCGL1to1}, viz., $\alpha=-1.5$, $\beta=6$ and $\mu=-1$, and vary the remaining two parameters $\nu$ and $\gamma$. Numerical continuation and DNS are, respectively, performed using AUTO-07p~\cite{auto} and the ETD2 scheme~\cite{CM02}. Unless otherwise specified, the domain size for DNS is $[-100,100]\times[-100,100]$.

\section{Circular fronts}\label{sec:circtw}

The properties of 1D fronts between $A^\pm$ in Eq.~(\ref{eq:FCGL1to1}) are well understood \cite{MBK10,MaKn12} following earlier work in \cite{KW:05}. In a bifurcation diagram of Type-I LS formed by two steady 1D fronts assembled back-to-back, the successive folds converge on a limiting parameter value in an oscillatory fashion as $A^+$ grows longer~\cite{MBK10}. This bifurcation behavior is termed {\it collapsed snaking}, and the limiting parameter value is termed the collapse point. At fixed $\nu=5$ the collapse point occurs at $\gamma^{CS}=1.8419$. Traveling 1D fronts at $\gamma\neq\gamma^{CS}$ correspond to heteroclinic orbits between $A^\pm$ of an ODE in the comoving frame~\cite{MaKn12}.

In this section we study a family of 2D circular fronts between $A^\pm$ in Eq.~(\ref{eq:FCGL1to1}). As in Refs.~\cite{MBK10,MaKn12} we fix $\nu=5$ and vary $\gamma$ as the bifurcation parameter. In \S\ref{sec:circfr-rcs} we compute the bifurcation diagram of steady circular fronts, and compare the observed bifurcation behavior with 1D collapsed snaking. In \S\ref{sec:circfr-spec} we compute the spectral stability of this solution branch, and demonstrate the existence of stable oscillons and the presence of different types of symmetry-breaking instabilities. In \S\ref{sec:circfr-tcf} we compute traveling circular fronts using both 2D and axisymmetric time evolution, and show that these correspond to heteroclinic orbits between $A^\pm$ in the radially comoving ODE.

\subsection{Radially collapsed snaking}\label{sec:circfr-rcs}

\begin{figure}
\center \subfigure[] {\label{fig:HEX_circtw_snake5} \mbox{\includegraphics[width=0.48\textwidth]{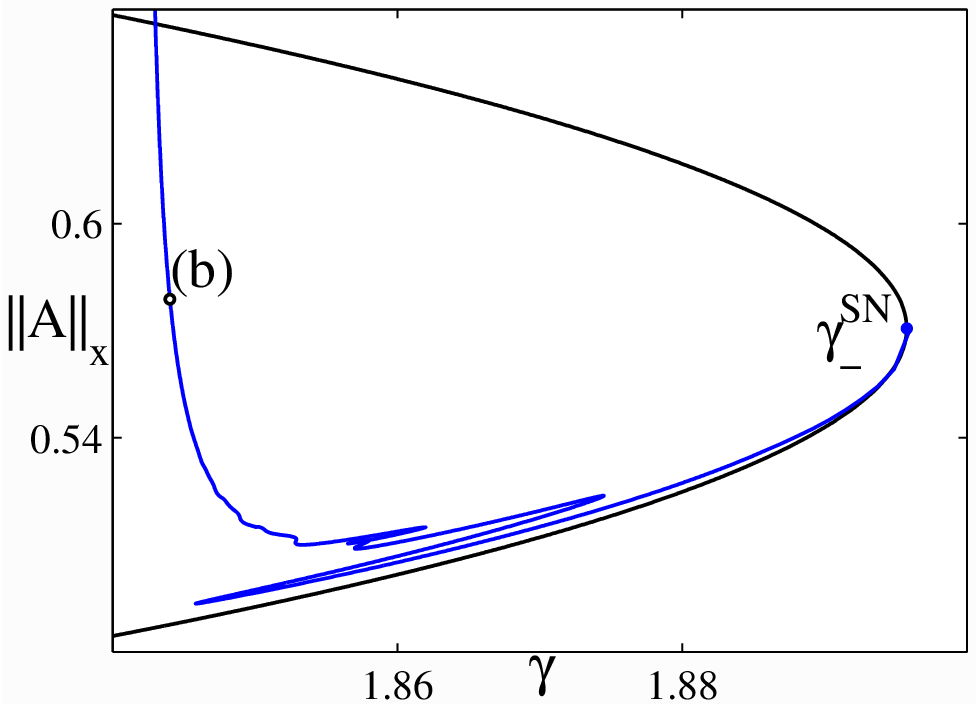}}} \subfigure[] {\label{fig:HEX_circfr_g1.844_dr1_t0} \mbox{\includegraphics[width=0.4\textwidth]{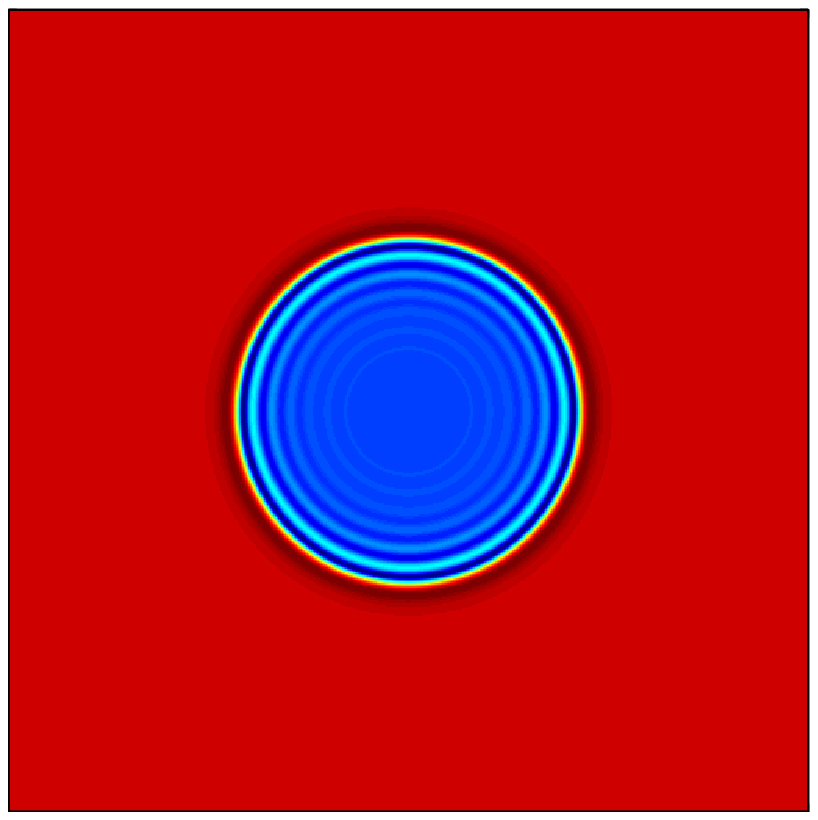}}} \caption{(a) The branch of 2D axisymmetric steady states followed from the lower saddle-node $\gamma_-^{SN}$ at $\nu=5$. (b) A sample solution profile $V(x,y)$ at $\gamma=1.844$.} \label{fig:HEX_circtw_steady}
\end{figure}

As pointed out in \S\ref{sec:intro}, steady $d$-dimensional axisymmetric solutions to Eq.~(\ref{eq:FCGL1to1}) satisfy Eq.~(\ref{eq:circfr_steadyn}). Since in 1D ($d=1$) the collapsed snaking branch of Type-I LS originates from the lower saddle-node $\gamma_-^{SN}$ of the S-shaped bifurcation diagram of equilibria (Fig. \ref{fig:HEX_elements}), we may postulate the same for the branch of steady circular fronts in 2D ($d=2$). In 1D weakly nonlinear solutions near $\gamma_-^{SN}$ can be derived using standard asymptotic methods (eg., Ref.~\cite{BYK:08}) and these can be continued from 1D to 2D by incrementing $d$ from $d=1$ to $d=2$ in Eq.~(\ref{eq:circfr_steadyn}); see Ref.~\cite{McQS14} for an analytical description of these 2D weakly nonlinear solutions in the 2:1 FCGLE. Once $d=2$ has been reached the resulting solutions can then be continued in $\gamma$ with $d=2$ fixed, namely within Eq.~(\ref{eq:circfr_steady2}). Figure~\ref{fig:HEX_circtw_snake5} shows the branch of steady 2D axisymmetric solutions thus obtained. As the branch is followed towards larger $L^2$-norm, the weakly nonlinear solutions homoclinic to $A^-$ do indeed grow into steady circular fronts between $A^\pm$ as postulated.

A circular front has a well-defined radius, $r=\rho$, which may be measured by
\begin{equation}\label{eq:rho-def}
V(r=\rho)=\frac{1}{2}(V^++V^-).
\end{equation}
As shown in Fig.~\ref{fig:HEX_circtw_snake5}, in the upper part the branch approaches the limit $\gamma=\gamma^{CS}$ as the norm $||A||_x$ or equivalently the front radius $\rho$ increases. This is because at $\rho=\infty$, the curvature term $r^{-1}\partial_r$ in Eq.~(\ref{eq:circfr_steady2}) vanishes and the steady circular front becomes identical to the steady 1D front, which exists only at $\gamma=\gamma^{CS}$. In general, the instantaneous speed $c$ of a circular front is a smooth function of $\gamma$ and $\rho$. Moreover, a circular front with large $\rho$ is well described by the governing ODE in $r$ restricted to an interval of width $dr$ around $r=\rho$ where $1\ll dr\ll\rho$. Hence for $\gamma$ near $\gamma^{CS}$ and $\rho$ large, the front speed $c$ should depend linearly on $d\gamma\equiv\gamma-\gamma^{CS}$ and the front curvature $\kappa\equiv\rho^{-1}$ as
\begin{equation}\label{eq:circfr-cgaro}
c=c_\gamma d\gamma+c_\kappa\kappa.
\end{equation}
The function $-c(d\gamma)$ at $\kappa=0$ is plotted in Fig.~3 in Ref.~\cite{MaKn12}, while $d\gamma(\kappa)$ at $c=0$ can be determined from Fig.~\ref{fig:HEX_circtw_snake5} of this paper using Eq.~(\ref{eq:norm-rho}). From the linearization of these two functions around the origin, we obtain numerically
\begin{equation}
c_\gamma=11.86,\quad c_\kappa=-1.113.
\end{equation}
In Appendix~\ref{app:circfr-var} we derive the analog of Eq.~(\ref{eq:circfr-cgaro}) analytically for variational PDEs.

\begin{figure}
\centering \includegraphics[width=0.48\textwidth]{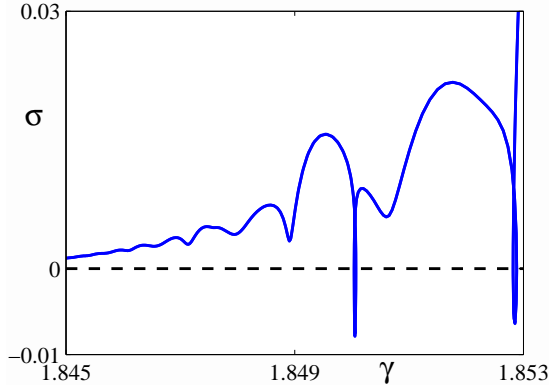}
\caption{The growth rate $\sigma$ of the amplitude mode $A_a(r)$ as a function of $\gamma$ along the branch of steady circular fronts in Fig.~\ref{fig:HEX_circtw_snake5}, computed using numerical continuation.} \label{fig:HEX_circfr_siga5}
\end{figure}

Using the definition $c=d\rho/dt$, we can rewrite Eq.~(\ref{eq:circfr-cgaro}) as the following ODE for $\kappa$:
\begin{equation}\label{eq:circfr-ode-kappa}
d\kappa/dt=-\kappa^2(c_\gamma d\gamma+c_\kappa\kappa).
\end{equation}
The nontrivial equilibrium and its eigenvalue within Eq.~(\ref{eq:circfr-ode-kappa}), denoted by
\begin{equation}\label{eq:circfr-kappa0-sigma0}
\kappa_0=-c_\gamma d\gamma/c_\kappa,\quad\sigma_{\kappa_0}=-c_\kappa\kappa_0^2=-c_\gamma^2(d\gamma)^2/c_\kappa,
\end{equation}
correspond, respectively, to the steady circular front $A_0(r)$ and an eigenvalue in the $\mathbb{D}_0$ spectrum of $A_0(r)$. As stated above, $A_0(r)$ becomes a steady 1D front located at $\rho_0\equiv\kappa_0^{-1}\rightarrow\infty$ as $d\gamma\rightarrow0$. In this limit, in the $\mathbb{D}_0$ spectrum of $A_0(r)$ the eigenvalue $0$ is the only marginal eigenvalue and has multiplicity $1$. The corresponding eigenfunction is the amplitude mode $A_a(r)=A_0'(r)$, which coincides with the Goldstone mode resulting from the invariance of Eq.~(\ref{eq:circfr_steady2}) under translation in $r$ as $r\rightarrow\infty$. As $d\gamma$ increases, the radius $\rho_0$ of a stationary front decreases and the translational invariance is broken more and more strongly. As a result, the amplitude (Goldstone) mode acquires a nonzero growth rate as given by Eq.~(\ref{eq:circfr-kappa0-sigma0}). As shown in Fig.~\ref{fig:HEX_circfr_siga5}, the growth rate $\sigma$ of the amplitude mode $A_a(r)$ along the branch determined by continuation does indeed agree with Eq.~(\ref{eq:circfr-kappa0-sigma0}) for small $d\gamma$.

As shown in Fig.~\ref{fig:HEX_circtw_snake5}, the LS branch first undergoes a series of ten folds, which we label as $\gamma_i$, $i=1,\cdots,10$, counting from $\gamma_-^{SN}$. The distance between successive folds increases non-monotonically as the oscillatory tail around $A^+$ is annihilated at $r=0$. This behavior is reminiscent of 1D collapsed snaking (CS), and will therefore be referred to as radially collapsed snaking (RCS). The oscillatory tail around $A^+$ results from the fact that the spatial eigenvalues of $A^+$ form a complex quartet denoted by $(\lambda_r\pm i\lambda_i,-\lambda_r\pm i\lambda_i)$ with $\lambda_{r,i}>0$. For CS, it was shown in Ref.~\cite{KW:05} (Lemma 2.3) that as the front radius $\rho$ increases, $\gamma$ approaches the limiting value $\gamma^{CS}$ according to
\begin{equation}\label{eq:circfr-cs-scaling}
\gamma-\gamma^{CS}\sim\exp{(-2\lambda_r\rho)}\sin{(2\lambda_i\rho+\phi)},
\end{equation}
where $\phi$ is a constant. For $d$-dimensional axisymmetric fronts in variational PDEs, it is shown in Appendix~\ref{app:circfr-var} that if either $\lambda_r\ll1$ or $d-1\ll1$, then RCS results from a competition between the $O(\exp(-2\lambda_r\rho))$ tail oscillation consistent with Eq.~(\ref{eq:circfr-cs-scaling}) and the monotonic contribution $O((d-1)\rho^{-1})$ proportional to the front curvature. If this also holds for the 1:1 FCGLE, then as $\rho$ decreases from $\infty$, folds should begin to appear at $\rho=\rho^{RCS}\gg1$, where $\rho^{RCS}$ is given by Eq.~(\ref{eq:var-RCS-eq}) of Appendix~\ref{app:circfr-var}. In the $\lambda_r\ll1$ case, this equation can be solved asymptotically to yield
\begin{equation}\label{eq:circfr-rho-RCS}
\rho^{RCS}=\lambda_r^{-1}(-\log{\lambda_r}+\log{(-\log{\lambda_r})}+O(1)).
\end{equation}
In the present parameter regime, for $\gamma$ not much larger than $\gamma^{CS}$, the spatial eigenvalues are $\lambda_r\approx0.19$ and $\lambda_i\approx1.4$. Hence the predicted critical radius using Eq.~(\ref{eq:circfr-rho-RCS}) is $\rho^{RCS}\approx8.8$, while the wavelength of the oscillatory tail is $2\pi/\lambda_i\approx4.5$. In comparison, the measured radii at $\gamma_{10}$ and $\gamma_8$ on the branch are respectively $\rho_{10}\approx10.2$ and $\rho_8\approx7.5$. As shown in Fig.~\ref{fig:HEX_circfr_siga5}, the growth rate $\sigma$ of the amplitude mode $A_a(r)$ changes sign at $\gamma_i$, $i=7,\cdots,10$. As a result, stable oscillons with radii approximately $\rho_{10}$ and $\rho_8$ exist, respectively, in the narrow intervals $(\gamma_9,\gamma_{10})$ and $(\gamma_7,\gamma_8)$, and no stable oscillon exists with radius greater than $\rho_{10}$. The existence of a maximum stable radius is a unique feature of axisymmetric oscillons in two or higher dimensions.

As the branch reaches $\gamma_6$ from above, the solution becomes well localized around $r=0$ with the front radius $\rho$ comparable to the tail wavelength. Thus the RCS mechanism based on the annihilation of the oscillatory tail cannot account for the lower folds. The complete spectral stability of the entire branch is computed directly in the next section.

\subsection{Spectral stability}\label{sec:circfr-spec}
\begin{figure}
\centering \includegraphics[width=0.96\textwidth]{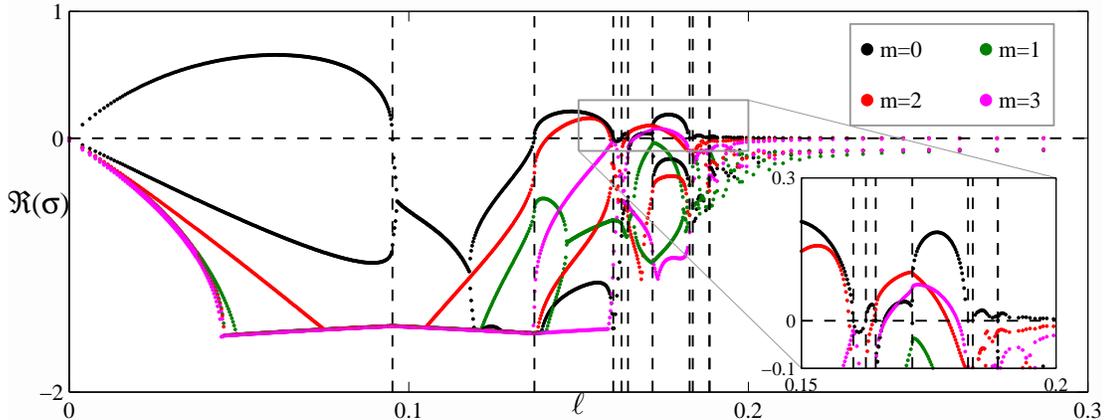}
\caption{Spectral stability of steady axisymmetric oscillons on the RCS branch in Fig.~\ref{fig:HEX_circtw_snake5}, shown in terms of the real part of the eigenvalue $\Re(\sigma)$ as a function of the arclength $\ell$ measured from $\gamma_-^{SN}$. For each azimuthal wavenumber $m=0,1,2,3$, only the two most unstable eigenvalues are plotted. The zero eigenvalue for $m=1$ resulting from translation invariance in space is not shown. The locations of the folds on the RCS branch are indicated using dashed vertical lines.} \label{fig:HEX_circfr_specrad5_azi}
\end{figure}

Figure~\ref{fig:HEX_circfr_specrad5_azi} shows the spectral stability of steady axisymmetric oscillons on the RCS branch in Fig.~\ref{fig:HEX_circtw_snake5}, where $\Re(\sigma)$ is the real part of the eigenvalue $\sigma$, and $\ell$ is the arclength along the branch relative to the bifurcation point at $\gamma_-^{SN}$. The $\mathbb{D}_m$ spectra are plotted only for the lowest azimuthal wavenumbers $m=0$ to $3$ since oscillons with small radii are most likely unstable to $\mathbb{D}_m$ perturbations with small $m$. For each $m$, only the two largest values of $\Re(\sigma)$, or equivalently the two most unstable or leading eigenvalues, are plotted. Thus at any $\ell$, the presence of two values of $\Re(\sigma)$ implies that the leading eigenvalue is real, whereas a single value of $\Re(\sigma)$ implies that the leading eigenvalues form a complex conjugate pair. Owing to translation invariance there is always a zero eigenvalue in the $\mathbb{D}_1$ spectrum, which is filtered out before the spectrum is plotted. The locations of the folds on the RCS branch are labeled using dashed vertical lines.

The $\mathbb{D}_0$ spectrum is shown in black in Fig.~\ref{fig:HEX_circfr_specrad5_azi}. There is an amplitude mode $A_a(r)$ whose eigenvalue is real and changes sign at every fold. As a result, oscillons are $\mathbb{D}_0$ unstable on the segments $\gamma_-^{SN}\rightarrow\gamma_1$, $\gamma_2\rightarrow\gamma_3$, and $\gamma_4\rightarrow\gamma_5$, and $\mathbb{D}_0$ stable on the segments $\gamma_1\rightarrow\gamma_2$ and $\gamma_3\rightarrow\gamma_4$. Interestingly, on the segment $\gamma_1\rightarrow\gamma_2$, the real eigenvalue of the amplitude mode $A_a(r)$ and a smaller real eigenvalue first coalesce into a complex conjugate pair, and this pair then splits back into the real eigenvalue of the amplitude mode $A_a(r)$ and a smaller real eigenvalue. Thus, for oscillons on this segment, $\mathbb{D}_0$ perturbations may decay in either monotonic or oscillatory fashion depending on $\gamma$. This process of coalescence and splitting happens again on the segment $\gamma_5\rightarrow\gamma_6$, but this time the complex conjugate pair resulting from the coalescence of the two real eigenvalues becomes unstable before splitting back to two real eigenvalues, this time with the smaller eigenvalue corresponding to the amplitude mode $A_a(r)$. Thus, for oscillons on this segment, $\mathbb{D}_0$ perturbations may grow or decay depending on $\gamma$ and may do so in either monotonic or oscillatory fashion. Beyond $\gamma_6$, oscillons are $\mathbb{D}_0$ unstable on the segments $\gamma_6\rightarrow\gamma_7$, $\gamma_8\rightarrow\gamma_9$, and $\gamma_{10}\rightarrow\gamma^{CS}$, and $\mathbb{D}_0$ stable on the segments $\gamma_7\rightarrow\gamma_8$ and $\gamma_9\rightarrow\gamma_{10}$, as explained by the RCS mechanism in \S\ref{sec:circfr-rcs}.

The $\mathbb{D}_m$ spectra for $m=1,2,3$ are shown, respectively, in green, red, and magenta. Since the odd segments $\gamma_{2i}\rightarrow\gamma_{2i+1}$, $i=0,1,\cdots$, with $\gamma_0\equiv\gamma_-^{SN}$ are always $\mathbb{D}_0$ unstable, we only consider $\mathbb{D}_m$ stability, $m\geq1$, on the even segments $\gamma_{2i+1}\rightarrow\gamma_{2i+2}$. We observe that the leading $\mathbb{D}_1$ eigenvalues are always stable, and are generally more stable than the leading $\mathbb{D}_2$ eigenvalues. The leading $\mathbb{D}_2$ and $\mathbb{D}_3$ eigenvalues are stable on all even segments except $\gamma_5\rightarrow\gamma_6$; on part of this segment, these eigenvalues are real and unstable.

In summary, oscillons on the even segments with relatively small or large radii, specifically those on $\gamma_{2i+1}\rightarrow\gamma_{2i+2}$, $i=0,1,3,4$, are spectrally stable. However, oscillons with intermediate radii, specifically those on $\gamma_5\rightarrow\gamma_6$, are always $\mathbb{D}_2$ unstable and may in addition be $\mathbb{D}_0$ and $\mathbb{D}_3$ unstable.

\begin{figure}
\center
\begin{tabular}{cccc}
\includegraphics[width=0.12\textwidth]{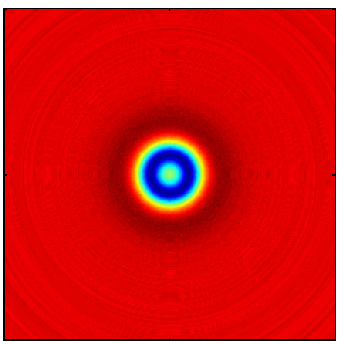} &
\includegraphics[width=0.12\textwidth]{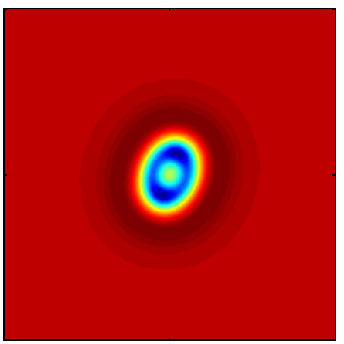} &
\includegraphics[width=0.12\textwidth]{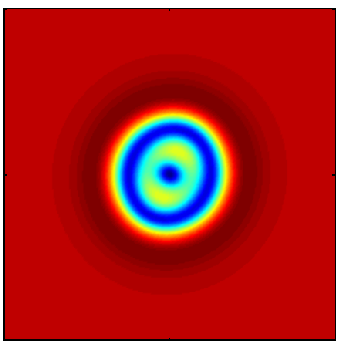} &
\includegraphics[width=0.12\textwidth]{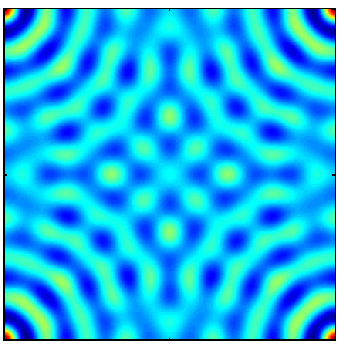}
\end{tabular}
\caption{Time evolution of a steady oscillon at $\gamma=1.857$ on the segment $\gamma_5\rightarrow\gamma_6$, with a $\mathbb{D}_2$ noise in the initial condition. In this case the only unstable eigenvalue is $0.050$ ($\mathbb{D}_2$). The domain size is $[-20,20]\times[-20,20]$, and the snapshots are taken at $t=0,100,200,400$.} \label{fig:HEX_nu5_aziosci_D2}
\end{figure}

\begin{figure}
\center
\begin{tabular}{cccccc}
\includegraphics[width=0.12\textwidth]{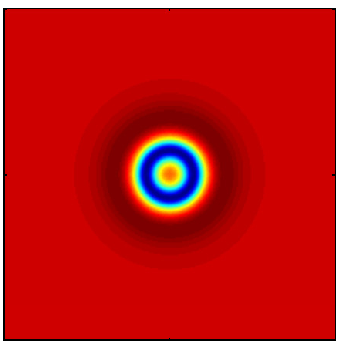} &
\includegraphics[width=0.12\textwidth]{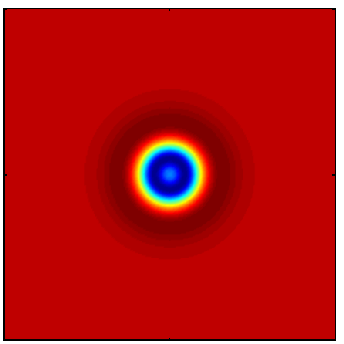} &
\includegraphics[width=0.12\textwidth]{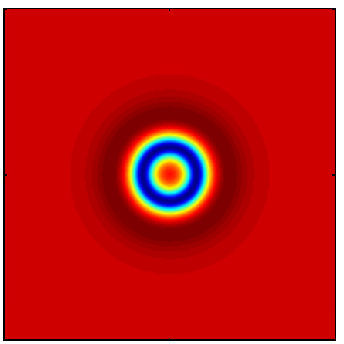} &
\includegraphics[width=0.12\textwidth]{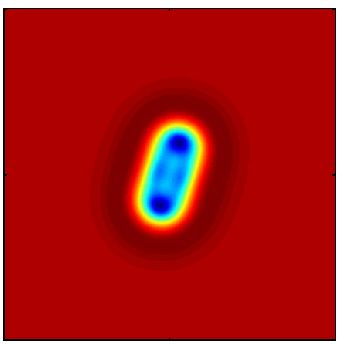} &
\includegraphics[width=0.12\textwidth]{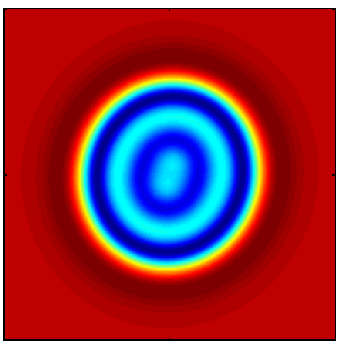} &
\includegraphics[width=0.12\textwidth]{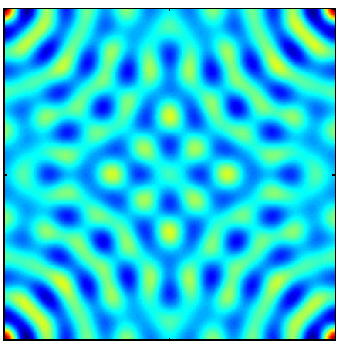}
\end{tabular}
\caption{Time evolution of a steady oscillon at $\gamma=1.86$ on the segment $\gamma_5\rightarrow\gamma_6$, with a $\mathbb{D}_0$ noise in the initial condition. In this case the unstable eigenvalues are $0.037\pm0.16i$ ($\mathbb{D}_0$), $0.092$ ($\mathbb{D}_2$), and $0.036$ ($\mathbb{D}_3$). The domain size is $[-20,20]\times[-20,20]$, and the snapshots are taken at $t=0,320,340,480,540,720$.} \label{fig:HEX_nu5_aziosci_D0D2}
\end{figure}

These instabilities may be observed by time-evolving steady oscillons with superposed small amplitude random noise. Figure~\ref{fig:HEX_nu5_aziosci_D2} shows the time evolution at $\gamma=1.857$ on the segment $\gamma_5\rightarrow\gamma_6$, where the only unstable eigenvalue is $0.050$ ($\mathbb{D}_2$). In this case, the $\mathbb{D}_2$ instability is triggered by the $\mathbb{D}_2$ noise in the initial condition, and the solution subsequently expands $\mathbb{D}_2$ symmetrically. After colliding with its images relative to the periodic domain boundaries, the solution first evolves into a spot homoclinic to the equilibrium $A^+$; this spot then persists for a long time before eventually collapsing into $A^+$ itself. Figure~\ref{fig:HEX_nu5_aziosci_D0D2} shows the time evolution at $\gamma=1.86$ on the segment $\gamma_5\rightarrow\gamma_6$, where the unstable eigenvalues are $0.037\pm0.16i$ ($\mathbb{D}_0$), $0.092$ ($\mathbb{D}_2$), and $0.036$ ($\mathbb{D}_3$). In this case, the noise triggers a $\mathbb{D}_0$ instability, and the solution subsequently oscillates $\mathbb{D}_0$ symmetrically. As the oscillation amplitude increases, the $\mathbb{D}_2$ instability takes over and the subsequent evolution beginning with a $\mathbb{D}_2$ symmetric expansion is similar to Fig.~\ref{fig:HEX_nu5_aziosci_D2}.

\subsection{Traveling circular fronts}\label{sec:circfr-tcf}
\begin{figure}
\center
\subfigure[] {\label{fig:HEX_circfr_g1.844_dr1_t9300} \mbox{\includegraphics[width=0.4\textwidth]{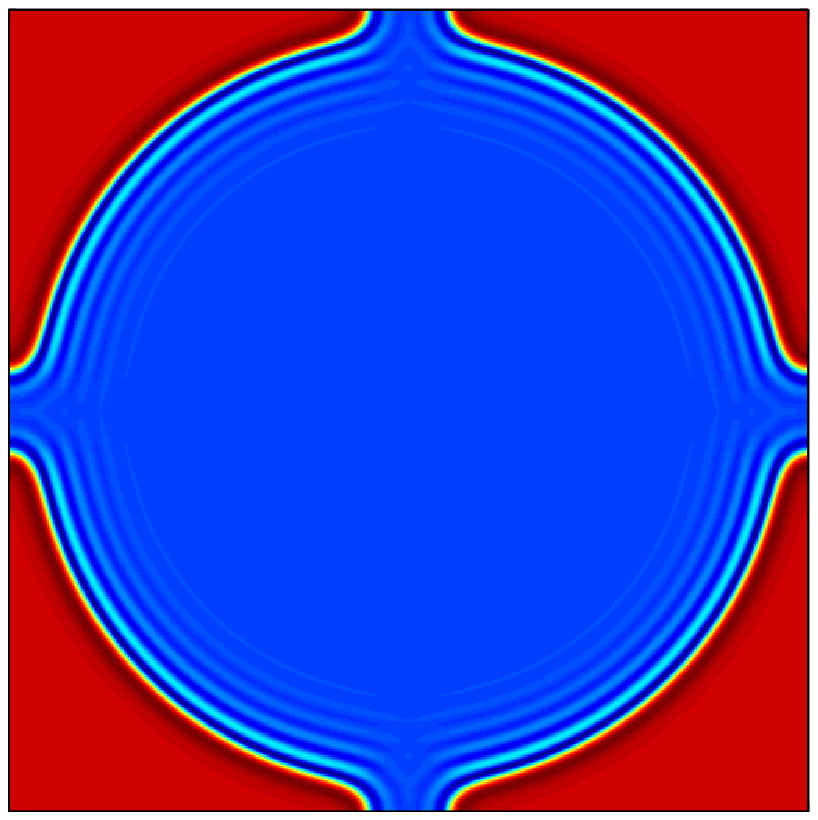}}}
\subfigure[] {\label{fig:HEX_circfr_g1.844_dr1_t9700} \mbox{\includegraphics[width=0.4\textwidth]{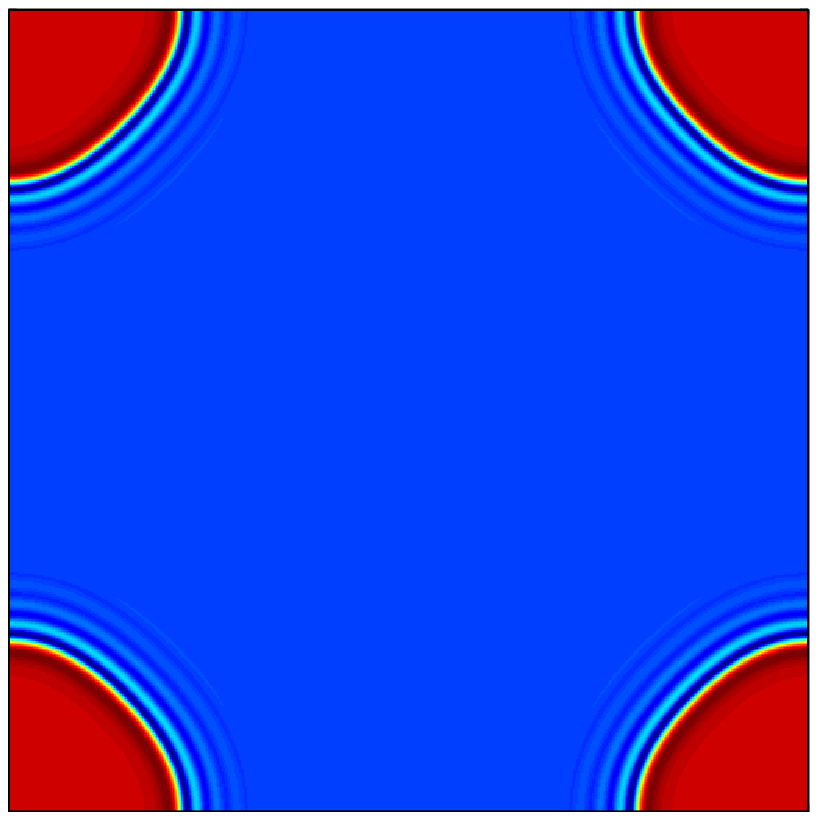}}}
\caption{Snapshots of $V(x,y)$ for $\nu=5$, $\gamma=1.844$, $d\gamma=0$, $d\rho=1$ at (a) $t=9300$ and (b) $t=9700$. The initial condition is the steady circular front in Fig.~\ref{fig:HEX_circfr_g1.844_dr1_t0}.} \label{fig:HEX_circtw_traveling}
\end{figure}

As shown in \S\ref{sec:circfr-rcs}, the radius $\rho$ of a steady circular front is determined locally uniquely by the value of the parameter $\gamma$. Traveling circular fronts are therefore expected when $\rho$ is changed by $d\rho$ or $\gamma$ is changed by $d\gamma$. When $d\rho<0$ ($d\rho>0$) and $d\gamma=0$, namely, when the front radius is contracted (expanded) at fixed $\gamma$, DNS shows that the circular front continues to contract (expand). Figure~\ref{fig:HEX_circtw_traveling} shows that in the latter case the circular front eventually collides with four image fronts, and evolves into reciprocal circular fronts centered at the corners, with $A^-$ now embedded in $A^+$. The solution eventually becomes spatially uniform when the $A^-$ bubbles in the corners contract to zero.

\begin{figure}
\center
\subfigure[] {\label{fig:HEX_circfr_g1.844_dr-1} \mbox{\includegraphics[width=0.4\textwidth]{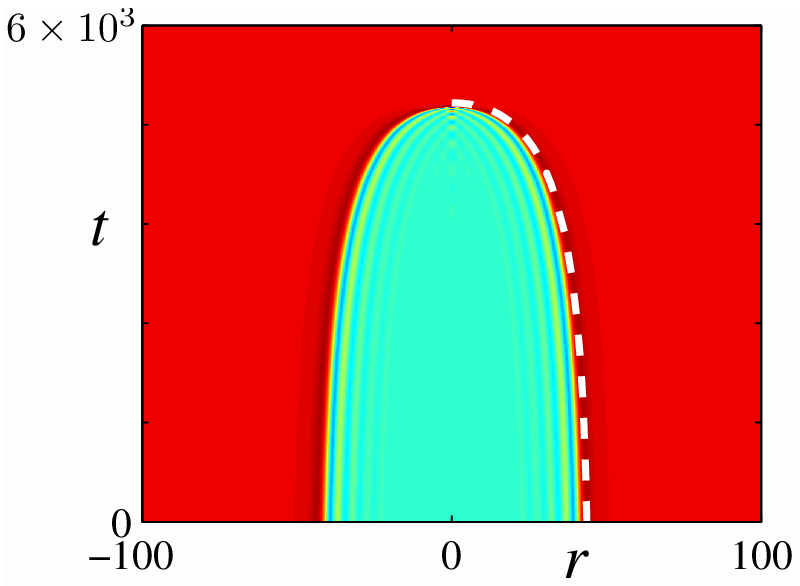}}}
\subfigure[] {\label{fig:HEX_circfr_g1.844_dr1} \mbox{\includegraphics[width=0.4\textwidth]{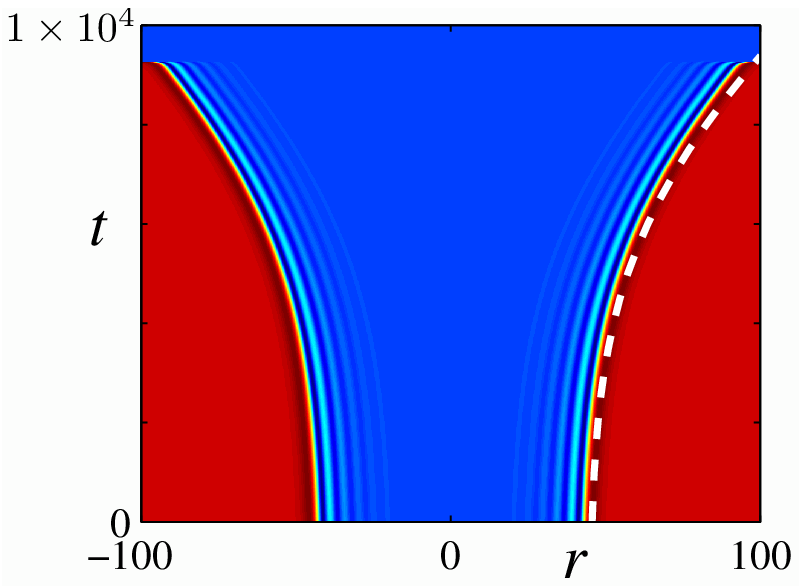}}}
\caption{Radial space-time plots of $V(r,t)$ for the same parameter values as in Fig.~\ref{fig:HEX_circtw_traveling}, except that (a) $d\rho=-1$, (b) $d\rho=1$. The dashed lines show solutions computed from Eq.~(\ref{eq:circfr-ode-kappa}) for comparison.} \label{fig:HEX_circtw_radial_only}
\end{figure}

The contraction and expansion processes maintain axisymmetry provided the LS does not reach the domain boundary. In general, axisymmetric time evolution is described by Eq.~(\ref{eq:FCGL_polar}) with $\partial_{\theta\theta}=0$, namely
\begin{equation}\label{eq:FCGL_radial_only}
A_t=(1+i\alpha)\left(\partial_{rr}+\frac{1}{r}\partial_r\right)A+f(A).
\end{equation}
Figure~\ref{fig:HEX_circtw_radial_only} shows radial space-time plots illustrating these processes computed from Eq.~(\ref{eq:FCGL_radial_only}), with the front evolution predicted using Eq.~(\ref{eq:circfr-ode-kappa}) shown as the dashed line. The agreement is excellent. In both contraction and expansion, we observe that the front speed increases as the front radius evolves away from its equilibrium value. The resulting front evolution is well approximated by Eq.~(\ref{eq:circfr-ode-kappa}) as long as the front radius $\rho$ remains large, or equivalently the front curvature $\kappa=\rho^{-1}$ remains small.

\begin{figure}
\centering \includegraphics[width=0.48\textwidth]{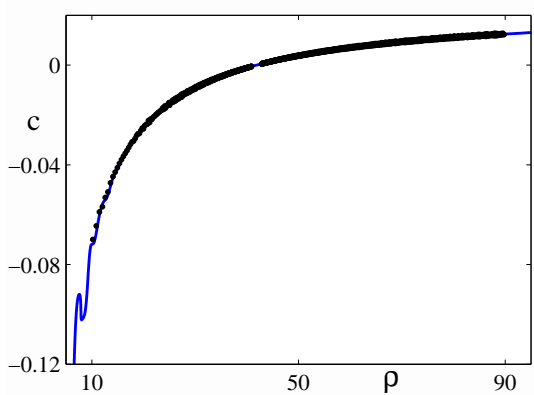}
\caption{The front speed $c$ as a function of the front radius $\rho$ at $\gamma=1.844$. Dotted: DNS. Solid: numerical continuation using Eq.~(\ref{eq:circfr_ode}). The curves superpose with high accuracy.} \label{fig:HEX_circfr_rv5}
\end{figure}

At any given time $t$ during the time evolution, the front radius $\rho$ can be measured by Eq.~(\ref{eq:rho-def}) and the instantaneous front speed $c$ can be computed as $c\equiv d\rho/dt$. Moreover, DNS results in Fig.~\ref{fig:HEX_circtw_radial_only} suggest that to a good approximation, $c$ is the instantaneous speed at which the entire front moves in the radial direction. Thus traveling circular fronts are well approximated by solutions to the radial PDE Eq.~(\ref{eq:FCGL_radial_only}) of the form $A(r-ct)$ near $t=0$. This ansatz turns Eq.~(\ref{eq:FCGL_radial_only}) into the radial traveling wave ODE
\begin{equation}\label{eq:circfr_ode}
-cA_r=(1+i\alpha)\left(\partial_{rr}+\frac{1}{r}\partial_r\right)A+f(A).
\end{equation}
Using this ODE with NBC given by Eq.~(\ref{eq:NBC}), the steady circular front in Fig.~\ref{fig:HEX_circfr_g1.844_dr1_t0} with $c=0$ can be numerically continued to obtain a family of traveling circular fronts with $c\neq0$. Figure~\ref{fig:HEX_circfr_rv5} shows the front speed $c$ as a function of the front radius $\rho$ measured from DNS and computed using continuation. The agreement is excellent for the range of $\rho$ over which $c$ is measured. In agreement with the RCS argument leading to Eq.~(\ref{eq:circfr-rho-RCS}), $c$ starts to oscillate as a function of $\rho$ near $\rho^{RCS}\approx8.8$.

\begin{figure}
\center
\begin{tabular}{cc}
\includegraphics[width=0.48\textwidth]{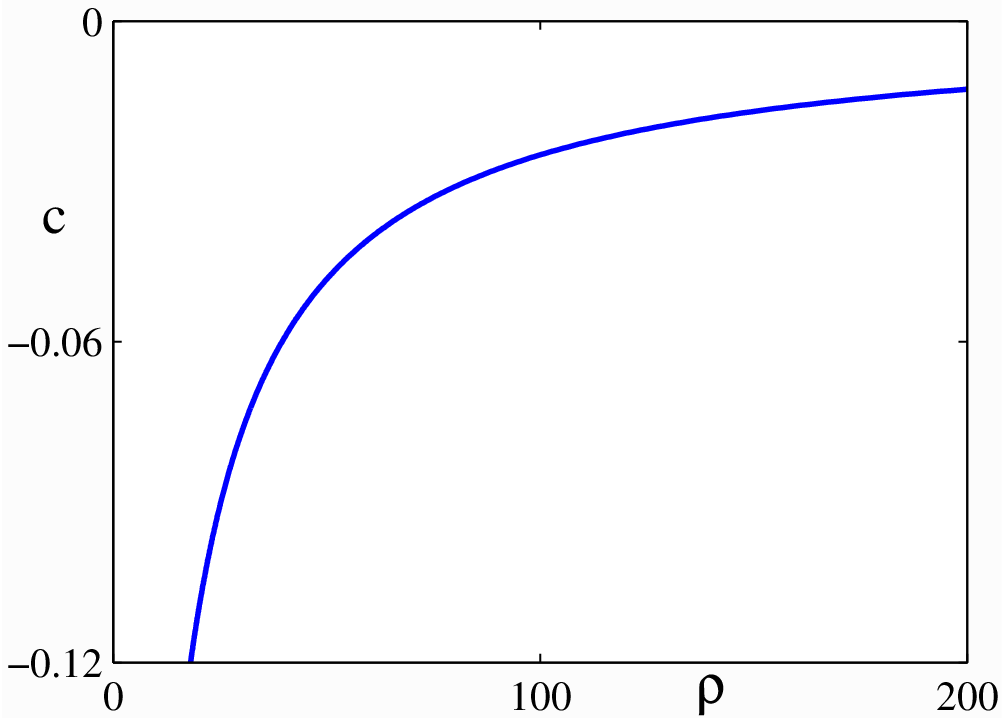} &
\includegraphics[width=0.4\textwidth]{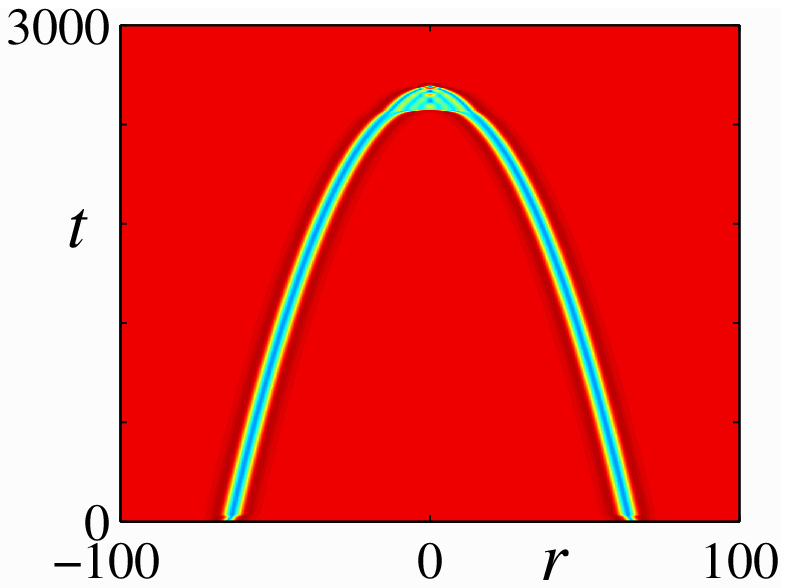} \\
(a) & (b)
\end{tabular}
\caption{A contracting pulse ring at $\gamma=1.844$. (a) The speed $c$ as a function of the radius $\rho$. (b) Radial space-time plot of $V(r,t)$ with initial condition corresponding to $c=-0.04$.} \label{fig:HEX_pulrin}
\end{figure}

Further continuation of this branch towards more negative front speed $c$ yields a family of contracting pulse rings. A contracting pulse ring is essentially a 1D stationary pulse bent into a circle in 2D, and thus its speed $c$ is proportional to its curvature $\kappa$; see Appendix~\ref{app:circfr-var} for an analytical description of contracting/expanding pulse rings in variational PDEs. As shown in Fig.~\ref{fig:HEX_pulrin}(a), the proportionality constant $c_\kappa\equiv c/\kappa$ is negative. The governing ODE for the ring radius, $\frac{d\rho}{dt}\equiv c=c_\kappa\kappa=\frac{c_\kappa}{\rho}$, can be solved to yield
\begin{equation}\label{eq:pulrin-frmo}
\frac{1}{2c_\kappa}\rho^2=t-t_0,
\end{equation}
with the constant $t_0$ determined by the initial condition. As shown in Fig.~\ref{fig:HEX_pulrin}(b), the contracting pulse ring taken from this branch is amplitude-unstable and thus rapidly evolves into a thicker contracting pulse ring. The latter contracts to zero in a parabolic fashion as predicted by Eq.~(\ref{eq:pulrin-frmo}) with $c_\kappa<0$; nontrivial dynamics are visible just prior to extinction when the radius of the ring becomes comparable to the pulse width. We remark that the evolution of a 1D stationary pulse bent into a closed 2D curve of an arbitrary shape may be analyzed following previous work on the Swift-Hohenberg equation~\cite{GlLi13}.

\section{Localized target patterns}\label{sec:locring}

The properties of time-independent Type-II LS of Eq.~(\ref{eq:FCGL1to1}) in 1D are explored in Refs.~\cite{MBK10,MaKn12}. These structures consist of a Turing pattern bifurcating from $A^+$ embedded in the homogeneous state $A^-$. The associated bifurcation diagram consists of a single solution branch oscillating between two distinct limiting parameter values; the oscillations reflect the growth of the Turing pattern via roll insertion at the center~\cite{MBK10}. This bifurcation behavior is termed defect-mediated snaking (DMS), the limiting parameter values are called the snaking limits, and the parameter interval inbetween is called the snaking interval. At fixed $\nu=7$ snaking takes place in the interval $[\gamma_1^{DMS},\gamma_2^{DMS}]=[2.8949,2.8970]$. Outside the snaking interval, Type-II LS depin by creating or destroying rolls in the interior of the wavetrain via successive phase slips~\cite{MaKn12}. These phase slips occur at the center of the LS when $\gamma$ is near either limit of the snaking region, but off-center for $\gamma$ farther away.

In this section we study a family of 2D localized target solutions of Eq.~(\ref{eq:FCGL1to1}) formed by embedding into $A^-$ an axisymmetric Turing pattern that bifurcates supercritically from $A^+$ (Fig. \ref{fig:HEX_elements}(a)). As in Refs.~\cite{MBK10,MaKn12} we fix $\nu=7$ and vary $\gamma$ as the bifurcation parameter. In \S\ref{sec:locring-cms} we compute the bifurcation diagram of steady localized target patterns, and compare their bifurcation behavior with 1D DMS and 2D RCS. In \S\ref{sec:locring-spec} we compute the spectral stability of solutions along this branch, and identify instabilities with possible relevance to experiments. In \S\ref{sec:locring-depin} we compute the radial depinning of localized target patterns, and compare their temporal dynamics with the depinning of Type-II LS in 1D.

\subsection{Core-mediated snaking}\label{sec:locring-cms}

\begin{figure}
\center
\subfigure[] {\label{fig:HEX_locring_snake7} \mbox{\includegraphics[width=0.48\textwidth]{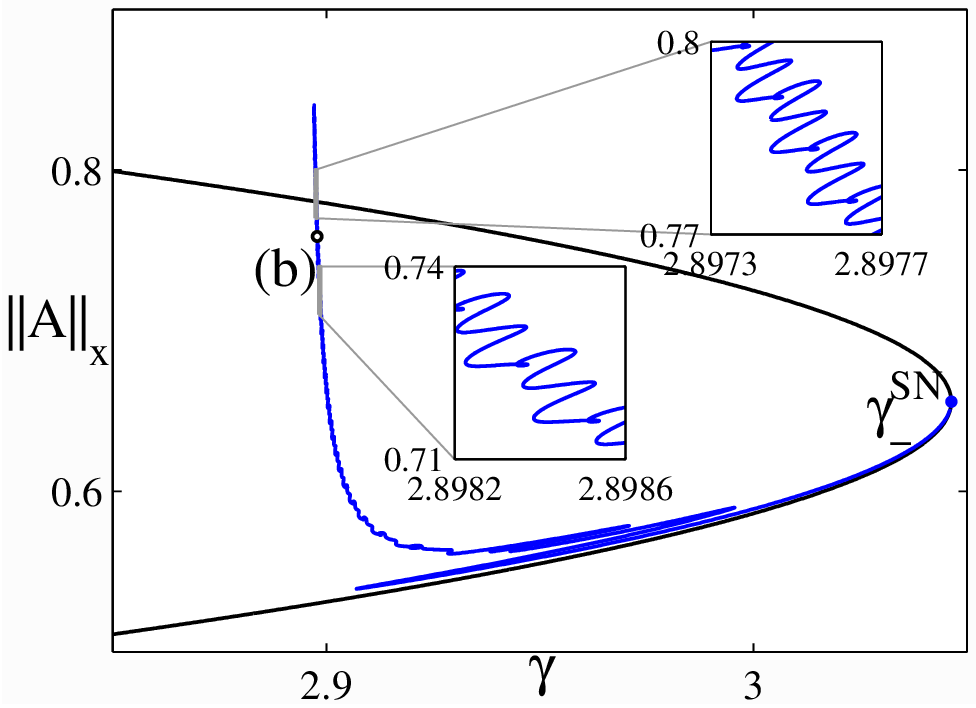}}}
\subfigure[] {\label{fig:HEX_circhex_g2.8989_t0} \mbox{\includegraphics[width=0.4\textwidth]{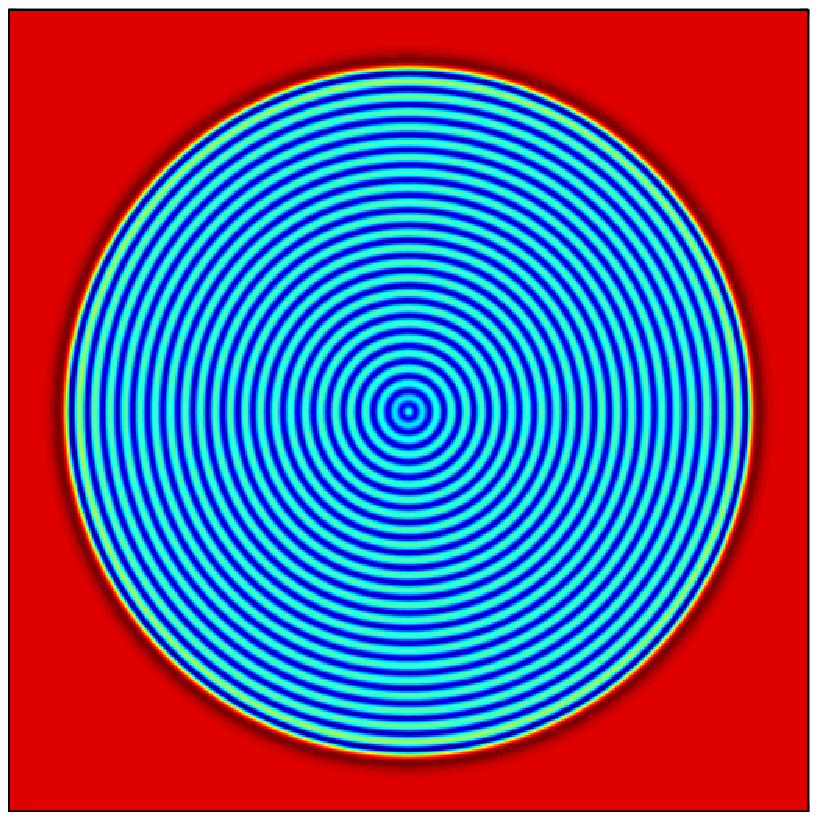}}}
\caption{(a) The branch of 2D axisymmetric steady states at $\nu=7$. A video showing the growth of the LS as the branch is followed from the lower saddle-node can be viewed at \href{\HEXlocringa}{this URL}. (b) A sample solution profile $V(x,y)$.} \label{fig:HEX_locring_steady}
\end{figure}

The branch of 2D axisymmetric steady states similar to Fig.~\ref{fig:HEX_circtw_snake5} but computed at $\nu=7$ is shown in Fig.~\ref{fig:HEX_locring_snake7}. Overall this branch is similar to the 2D RCS branch shown in Fig.~\ref{fig:HEX_circtw_snake5}, where the curvature of the circular front causes the upper part to converge monotonically to a single limiting parameter value. In contrast to Fig.~\ref{fig:HEX_circtw_snake5}, in the present case the branch undergoes $\xi$-shaped snaking on narrow intervals of $\gamma$. The separation between successive intervals decreases as the $L^2$-norm increases and the solutions high up the branch take the form of localized target patterns (Fig.~\ref{fig:HEX_circhex_g2.8989_t0}). As shown in the video accompanying Fig.~\ref{fig:HEX_locring_steady}, the localized target patterns in Eq.~(\ref{eq:FCGL1to1}) grow through a central defect at the core $r=0$, similar to 1D DMS. This bifurcation structure is hereafter referred to as core-mediated snaking (CMS).

The snaking mechanism of localized target patterns in the 1:1 FCGLE may be compared with the behavior of the corresponding structures in SH23~\cite{McCS10}. When $d=1$ new rolls in SH23 are nucleated at the location of the fronts while in the 1:1 FCGLE they are inserted by the division of the central ``cell'' or ``defect''. When $d>1$ localized target patterns in both equations possess a spike at $r=0$. As a result, as $d$ increases, the snaking mechanism in SH23 changes from 1D standard homoclinic snaking to 2D CMS via a stack of isolas. In contrast, in the 1:1 FCGLE the defect and the spike are both located at $r=0$ and as $d$ increases the snaking mechanism is expected to transition continuously from 1D DMS to 2D $\xi$-shaped CMS without the break-up of the snaking curve. We remark that in either SH23 or the 1:1 FCGLE, since the core plays the role of a ``pacemaker'' that emits new rings, the solutions with a local maximum/minimum at $r=0$ (known as type A/B solutions in Ref.~\cite{McCS10}) necessarily exist on the CMS branch in an alternative fashion.

\begin{figure}
\centering \includegraphics[width=0.48\textwidth]{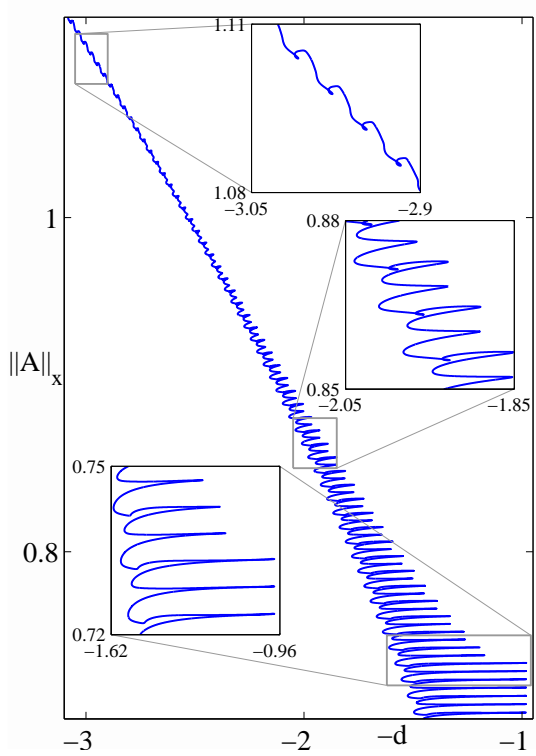}
\caption{The branch of $d$-dimensional axisymmetric steady states followed in $d$ from a 1D localized state ($d=1$) on the DMS branch at $\nu=7$, $\gamma=2.897$.} \label{fig:HEX_locring_dimcon}
\end{figure}

To test this conjecture, we have continued solutions to Eq.~(\ref{eq:circfr_steadyn}) in the dimension $d$; the result of this continuation at $\gamma=2.897$ is shown in Fig.~\ref{fig:HEX_locring_dimcon}. In order that the snaking in $d$ corresponds to snaking in $\gamma$ at nearby values of $d$, we plot the solution branch in the $(-d,\|A\|_x)$ plane, such that the sharper folds are located on the right. In this way a pair of adjacent C-shaped segments forms the body of the letter $\xi$. Near $d=1$ the snaking in $d$ takes place mostly in $d>1$ and resembles DMS except that the folds corresponding to maxima and minima in $V(r)$ at $r=0$ no longer align with each other. As $d$ increases a hysteresis loop appears on the bottom segment of every pair of adjacent C-shaped segments, while the snaking interval in $d$ becomes narrower. At $d=2$, the two additional segments forming this hysteresis loop can be regarded as the (short) tail of the current and the (long) head of the preceding letter $\xi$. Thus one extra ring is added at the core every time the branch passes through a $\xi$-shaped segment. At large $d$ (e.g.~$d=3$), the body of the $\xi$ disappears via two hysteresis bifurcations, although the head and tail remain. We have not continued the branch to even larger $d$ and thus cannot ascertain whether another hysteresis bifurcation would eventually lead to a bifurcation curve without folds, but from Fig.~\ref{fig:HEX_locring_dimcon} the branch appears to approach a straight line in the $(-d,\|A\|_x)$ plane as $d\rightarrow\infty$.

Apart from continuation in $d$ for $\gamma=2.897$ near $\gamma_2^{DMS}$, we have also examined other slices at constant $\gamma$ but these continuations do not lead to axisymmetric steady states in higher dimensions. Specifically, for smaller $\gamma$ (near $\gamma_1^{DMS}$) the branch snakes towards $d<1$ while for intermediate $\gamma$ (about halfway between $\gamma_1^{DMS}$ and $\gamma_2^{DMS}$) the branch snakes around $d=1$.

\subsection{Spectral stability}\label{sec:locring-spec}

\begin{figure}
\centering \includegraphics[width=0.96\textwidth]{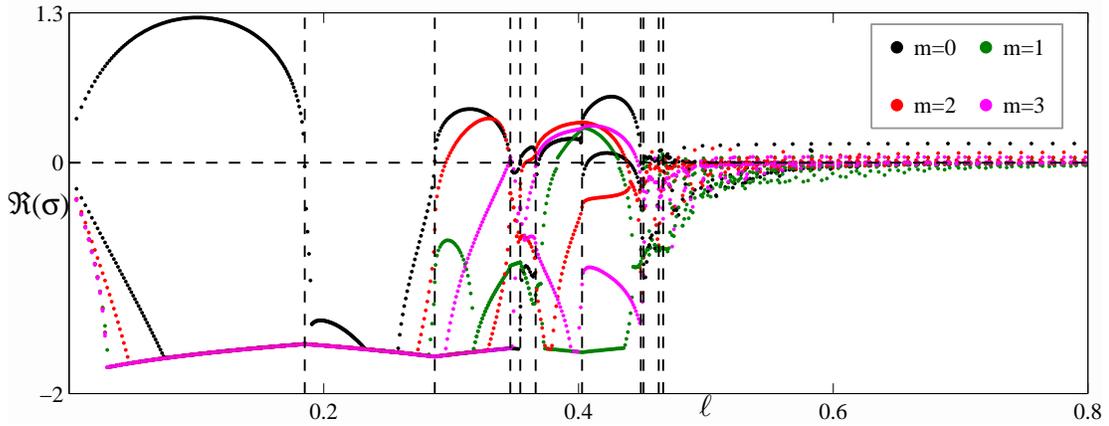}
\caption{Spectral stability of steady axisymmetric oscillons on the CMS branch in Fig.~\ref{fig:HEX_locring_snake7}, plotted using the same convention as in Fig.~\ref{fig:HEX_circfr_specrad5_azi}. For clarity only the first ten folds are indicated using dashed vertical lines.} \label{fig:HEX_locring_specrad5_azi}
\end{figure}

Figure~\ref{fig:HEX_locring_specrad5_azi} shows the spectral stability of steady axisymmetric oscillons on the CMS branch in Fig.~\ref{fig:HEX_locring_snake7}, plotted using the same conventions as in Fig.~\ref{fig:HEX_circfr_specrad5_azi}. As in Fig.~\ref{fig:HEX_circfr_specrad5_azi}, oscillons on even segments with relatively small radii are spectrally stable. In contrast to Fig.~\ref{fig:HEX_circfr_specrad5_azi}, oscillons with intermediate radii may be $\mathbb{D}_m$ unstable not only for $m=0,2,3$, but also for $m=1$. Also, oscillons with relatively large radii, which resemble localized target patterns, are generally $\mathbb{D}_m$ unstable for $m\geq1$ due to azimuthal instability of the rings.

\begin{figure}
\center
\begin{tabular}{cccccc}
\includegraphics[width=0.12\textwidth]{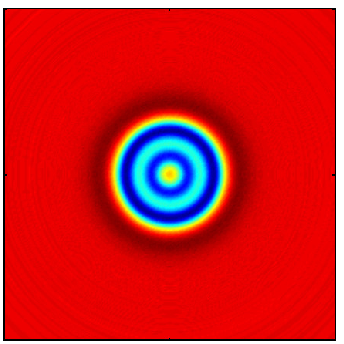} &
\includegraphics[width=0.12\textwidth]{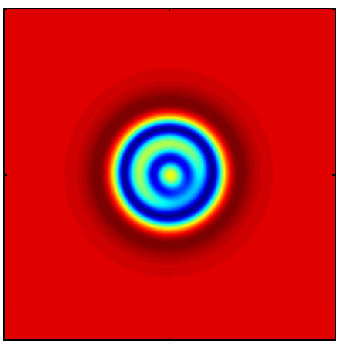} &
\includegraphics[width=0.12\textwidth]{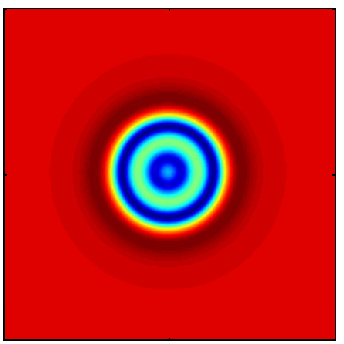} &
\includegraphics[width=0.12\textwidth]{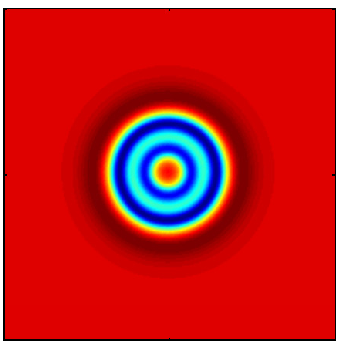} &
\includegraphics[width=0.12\textwidth]{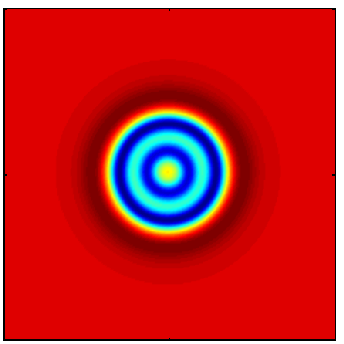} &
\includegraphics[width=0.12\textwidth]{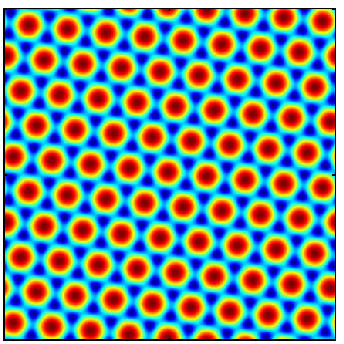}
\end{tabular}
\caption{Time evolution of a steady oscillon at $\gamma=2.922$ on the segment $\gamma_9\rightarrow\gamma_{10}$, with a $\mathbb{D}_1$ noise in the initial condition. In this case the unstable eigenvalues are $0.022\pm0.12i$ ($\mathbb{D}_0$) and $0.076$ ($\mathbb{D}_1$). The domain size is $[-20,20]\times[-20,20]$, and the snapshots are taken at $t=0,70,90,106,110,800$.} \label{fig:HEX_nu7_aziosci_D0D1}
\end{figure}

Figure~\ref{fig:HEX_nu7_aziosci_D0D1} shows the time evolution of a steady oscillon at $\gamma=2.922$ on the segment $\gamma_9\rightarrow\gamma_{10}$, where the unstable eigenvalues are $0.022\pm0.12i$ ($\mathbb{D}_0$) and $0.076$ ($\mathbb{D}_1$). In this case, the noise initially excites a $\mathbb{D}_1$ instability, and the solution subsequently develops a $\mathbb{D}_1$ symmetric interior. The $\mathbb{D}_0$ symmetry is subsequently restored, triggering the $\mathbb{D}_0$ instability. As a result, the solution first oscillates and then expands $\mathbb{D}_0$ symmetrically, before colliding with its images and eventually evolving into a hexagonal pattern.

For $m\geq1$, the locations on the RCS or CMS branch where the $\mathbb{D}_m$ stability changes correspond to bifurcation points to steady $m$-fold symmetric oscillons. Further from the onset, these $m$-fold symmetric oscillons may become stable, in which case they may be identified with the $m$-fold symmetric oscillon ``molecules'' observed in some experiments~\cite{UMS:96,LHARF:99}. Thus, dipoles formed by two out-of-phase oscillons may bifurcate from a $\mathbb{D}_1$ bifurcation point, dimers formed by two in-phase oscillons may bifurcate from a $\mathbb{D}_2$ bifurcation point, and so on. A detailed study of the interaction between simple oscillons and the bifurcation of oscillon molecules is left for future work.

\subsection{Radial depinning and traveling circular fronts}\label{sec:locring-depin}

\begin{figure}
\center
\subfigure[] {\label{fig:HEX_circfr_g2.897441_dg0_dr-1} \mbox{\includegraphics[width=0.48\textwidth]{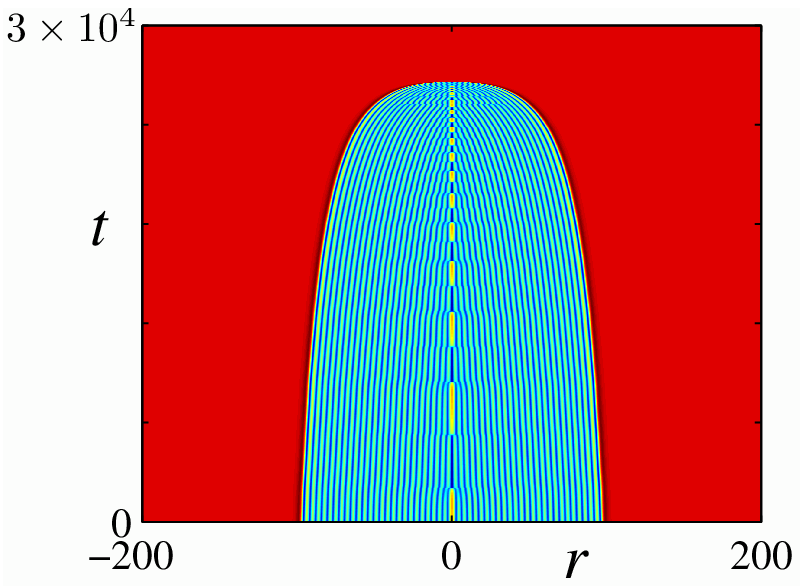}}}
\subfigure[] {\label{fig:HEX_circfr_g2.897441_dg-0.001_dr0} \mbox{\includegraphics[width=0.48\textwidth]{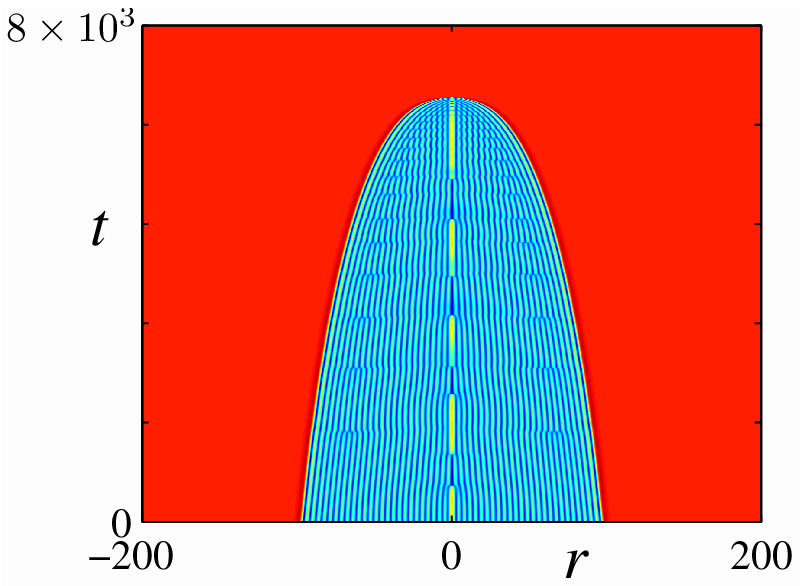}}}
\subfigure[] {\label{fig:HEX_circfr_g2.897556_dg0_dr1} \mbox{\includegraphics[width=0.48\textwidth]{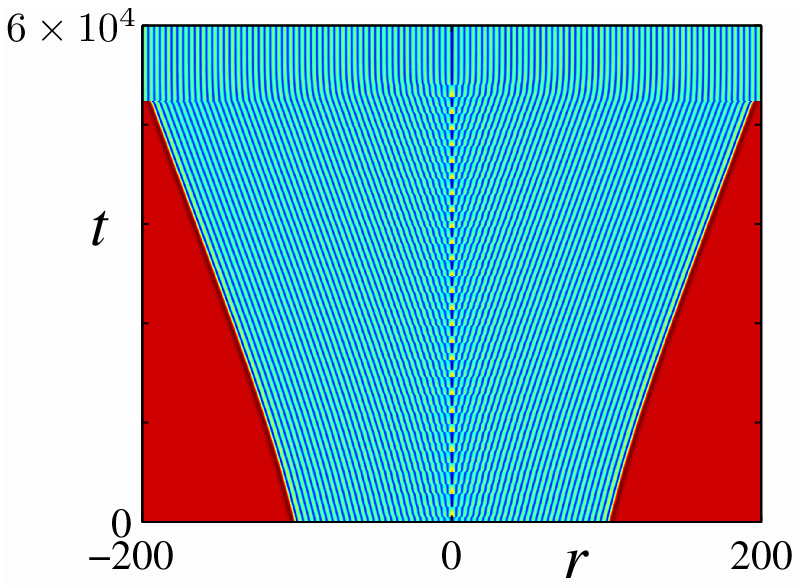}}}
\subfigure[] {\label{fig:HEX_circfr_g2.897556_dg0.002_dr0} \mbox{\includegraphics[width=0.48\textwidth]{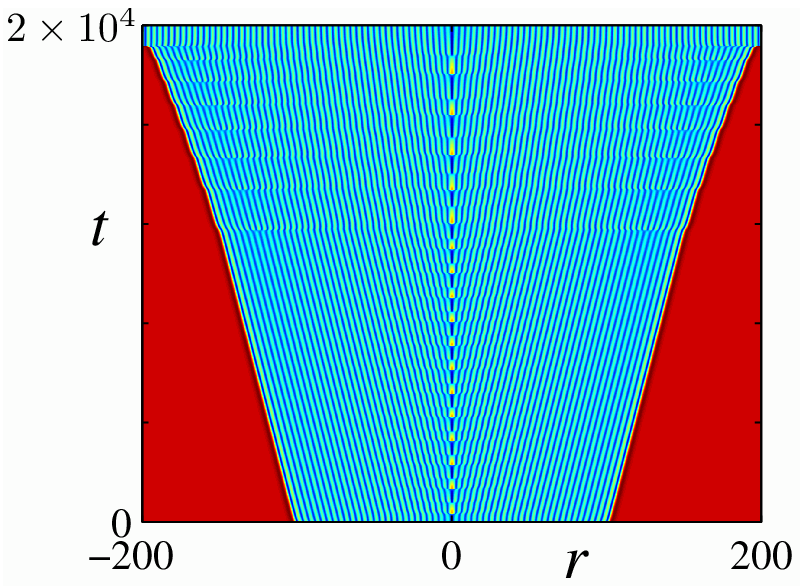}}}
\caption{Radial space-time plots of $V(r,t)$ at $\nu=7$ and near the fold at (a,b) $\gamma=2.897441$, (c,d) $\gamma=2.897556$, with the perturbations (a) $d\gamma=-1\times10^{-4}$; (b) $d\gamma=-1\times10^{-3}$; (c) $d\gamma=5\times10^{-4}$; (d) $d\gamma=2\times10^{-3}$.} \label{fig:HEX_locring_depin}
\end{figure}

Owing to the presence of azimuthal instabilities, fully 2D DNS of localized target patterns generally produce fully 2D localized states, as discussed further in \S\ref{sec:circhex}. On the other hand, azimuthal instabilities may be artificially suppressed by imposing axisymmetry on the solutions. This procedure allows us to time evolve localized target patterns using Eq.~(\ref{eq:FCGL_radial_only}) and to study radial depinning of the bounding front, as done in \S\ref{sec:circfr-tcf}. Since the front speed $c$ and the dynamically selected interior wavenumber $k$ are in general coupled \cite{MaKn12} we assume in the following that the wavenumber selection is largely passive (i.e., that $c$ depends only weakly on $k$). In this case the speed $c$ may be regarded as a function of the control parameter $\gamma$ and the front radius $\rho$ only, as in \S\ref{sec:circfr-tcf}. Hence to obtain $c\neq0$, we can start with a localized target pattern at a particular fold on the CMS branch in Fig.~\ref{fig:HEX_locring_snake7} and either change the radius $\rho$ by $d\rho$ or the parameter $\gamma$ by $d\gamma$; the outcome is similar. Figure~\ref{fig:HEX_locring_depin} shows radial space-time plots computed for different choices of the fold location and perturbation $d\gamma$. Figures~\ref{fig:HEX_circfr_g2.897441_dg0_dr-1} and \ref{fig:HEX_circfr_g2.897441_dg-0.001_dr0} show the time evolution near and far from the CMS region, respectively, in both cases starting from an initial condition at a left fold on the CMS branch; Figs.~\ref{fig:HEX_circfr_g2.897556_dg0_dr1} and \ref{fig:HEX_circfr_g2.897556_dg0.002_dr0} show the corresponding results for an initial condition at a right fold instead. In all four cases, the front speed increases as the front radius evolves away from its equilibrium value, and individual rings are destroyed or created in the interior of the structure via repeated phase slips. Thus, radial depinning of localized target patterns inherits the key features of both traveling circular fronts and depinning of Type-II LS in 1D. A particularly interesting new feature is that near the CMS region the phase slips occur right at the core, while far from the CMS region the phase slips occur both right at the core and at a nonzero radius away from the core.

Of particular interest is the behavior of the core at $r=0$. To the left of the left fold the structure contracts, but the way it does so depends on whether the perturbation is small in which case the frequency with which phase slips occur in the core accelerates as the target contracts (Fig.~\ref{fig:HEX_circfr_g2.897441_dg0_dr-1}), or the perturbation is large in which case the majority of the phase slips take place away from the core (Fig.~\ref{fig:HEX_circfr_g2.897441_dg-0.001_dr0}). Similar behavior is seen when the target expands. If the perturbation is small the phase slip frequency in the core continues to increase as the target expands (Fig.~\ref{fig:HEX_circfr_g2.897556_dg0_dr1}), but it decreases with time when the perturbation is large (Fig.~\ref{fig:HEX_circfr_g2.897556_dg0.002_dr0}). Thus the different magnitudes of the perturbation have distinct consequences. A small perturbation destabilizes the core ring first since curvature narrows the range of stable rings. The resulting phase slip eliminates a ring at the core but the inward front motion triggers additional phase slips at the core. As the structure contracts more and more rapidly the phase slip frequency necessarily increases. In contrast, when the structure contracts because of a large perturbation the inward motion of the front compresses mostly the outer rings (the curvature provides the inner rings with increased stiffness) leading to phase slips in the vicinity of the front. The converse happens when the structure expands. Thus the system ``remembers'' the initial perturbation that led to the contraction or expansion, in contrast to the 1D case in which the phase slips near the fronts on either side gradually migrate inwards to or outward from the center $x=0$ as the structure contracts or expands. We attribute these differences to the reduced role played by curvature in the outer regions of the pattern and hence to radial inhomogeneity of the contraction or expansion processes.

\begin{figure}
\center
\begin{tabular}{cc}
\includegraphics[width=0.48\textwidth]{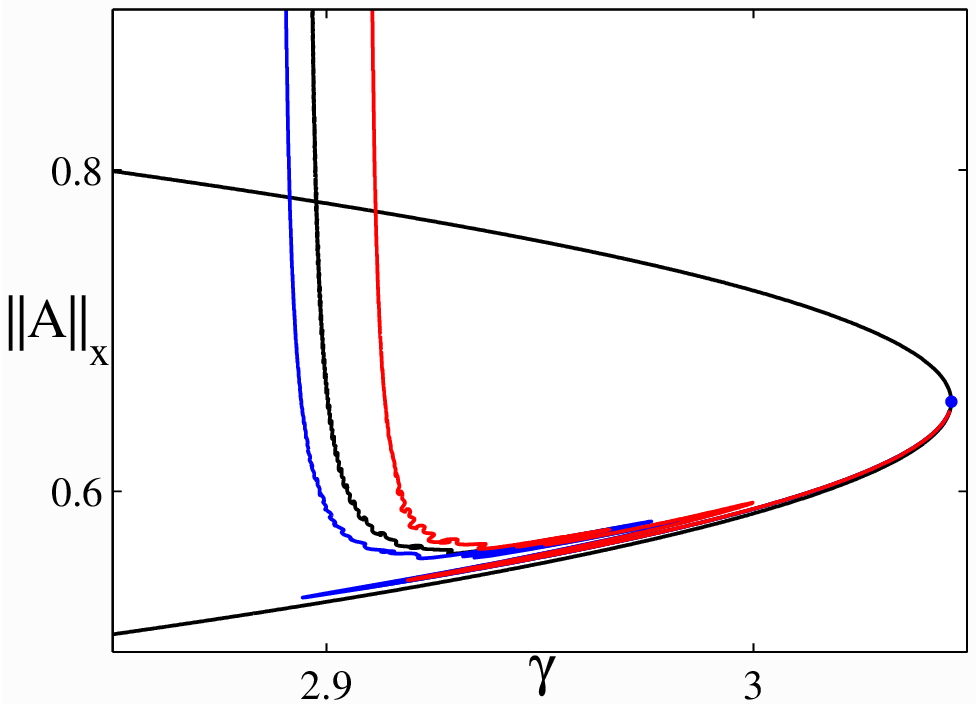} &
\includegraphics[width=0.48\textwidth]{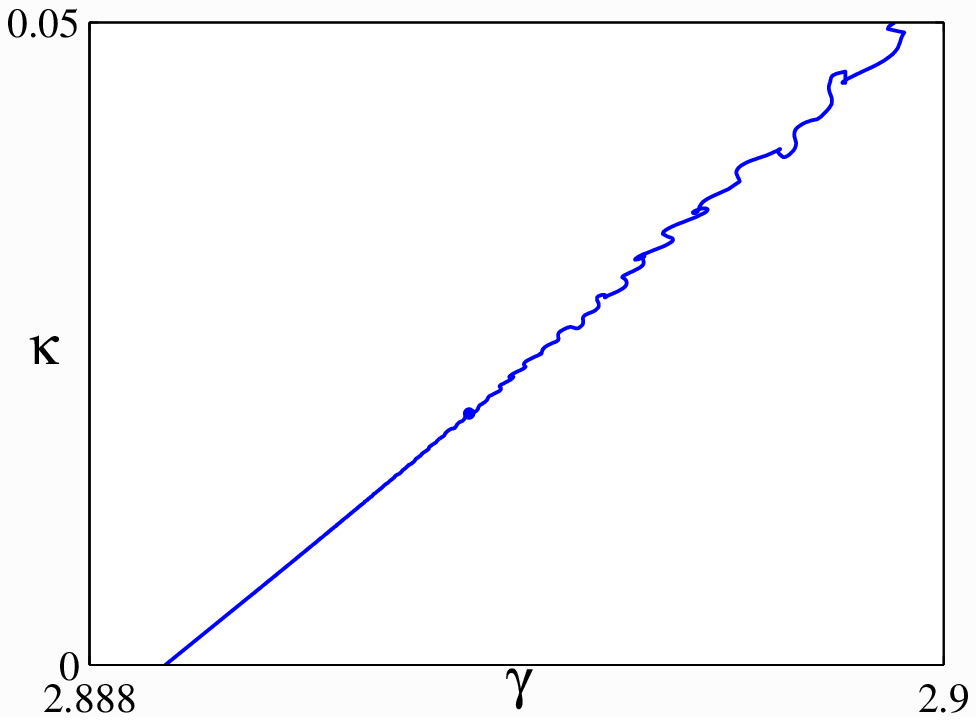} \\
(a) & (b)
\end{tabular}
\begin{tabular}{c}
\includegraphics[width=0.48\textwidth]{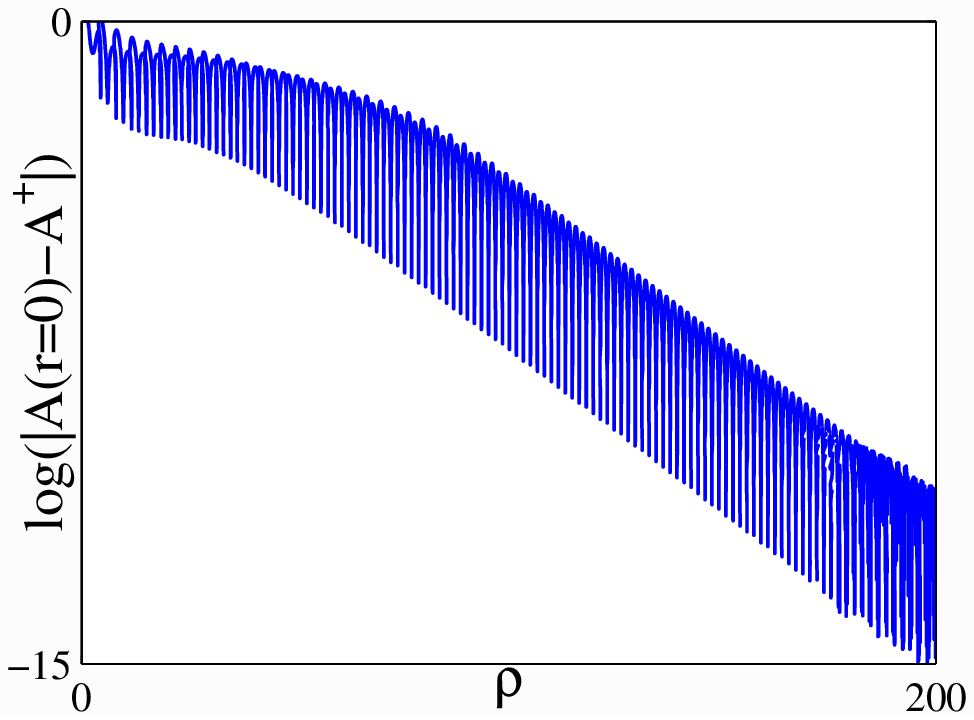} \\
(c)
\end{tabular}
\caption{(a) Two branches of traveling circular fronts computed at $c=-0.05$ (blue) and $c=0.05$ (red), and the $\xi$-shaped CMS branch of steady localized targets (black). (b) The bifurcation parameter $\gamma$ as a function of the curvature $\kappa$ at $c=-0.05$. (c) The logarithm of the core amplitude $|A(r=0)-A^+|$ as a function of the front radius $\rho$ at $c=-0.05$.} \label{fig:HEX_locring_tcf}
\end{figure}

As in the 1D case~\cite{MaKn12}, Eq.~(\ref{eq:circfr_ode}) contains codimension-1 traveling circular fronts between $A^\pm$ for either sign of the front speed $c$. Figure~\ref{fig:HEX_locring_tcf}(a) shows two branches of contracting/expanding circular fronts computed at $c=-0.05$ (blue) and $c=0.05$ (red), which are found to exist on opposite sides of the $\xi$-shaped CMS branch of steady localized target patterns (black). Near $c=0$, these two branches are expected to approach the left and right snaking limits of the CMS branch and thus remain separated by a finite distance. The origin of this phenomenon is the so-called ``reversibility-breaking'' bifurcation on $A^+$ at $c=0$~\cite{MaKn12}. As $c$ crosses zero, the two pairs of spatial eigenvalues $\pm ik_1$ and $\pm ik_2$ ($k_{1,2}>0$) of $A^+$ cross the imaginary axis simultaneously in opposite directions. As a result, the unstable eigendirections of $A^+$ change by a finite amount, and so does the value of $\gamma$ at which the heteroclinic orbit between $A^\pm$ exists.

For $0<|c|\ll1$, the real part $\lambda_r$ of the unstable eigenvalue of $A^+$ scales as $O(c)$~\cite{MaKn12}. Thus as discussed in \S\ref{sec:circfr-rcs}, the branch of traveling circular fronts between $A^\pm$ should exhibit RCS. Figure~\ref{fig:HEX_locring_tcf}(b) shows the bifurcation parameter $\gamma$ as a function of the curvature $\kappa$ at $c=-0.05$. The solid dot labels the critical radius $\rho^{RCS}=51$ beyond which folds disappear. At this point one calculates that $\lambda_r=0.087$, so the predicted critical radius based on Eq.~(\ref{eq:circfr-rho-RCS}) is $\rho^{RCS}=39$. Thus even for relatively small $\lambda_r$, Eq.~(\ref{eq:circfr-rho-RCS}) underestimates the critical radius $\rho^{RCS}$. To understand this discrepancy, we note that the underlying assumption of Eq.~(\ref{eq:circfr-rho-RCS}) that the tail oscillation is $O(\exp(-2\lambda_r\rho))$ presumes that the core amplitude is $O(\exp(-\lambda_r\rho))$. While this holds true for 1D LS, 2D axisymmetric LS are often found to have larger amplitude cores~\cite{Lloyd2,McCS10}. Figure~\ref{fig:HEX_locring_tcf}(c) shows the logarithm of the core amplitude $|A(r=0)-A^+|$ as a function of the front radius $\rho$ at $c=-0.05$. It can be seen that the decaying exponent is indeed $\lambda_r$ for large $\rho$, but becomes less than $\lambda_r$ for smaller $\rho$ including $\rho$ values around $\rho^{RCS}$.

\section{Localized hexagonal patterns}\label{sec:fully-2d}

In this section we study a family of 2D localized hexagonal patterns consisting of fully 2D hexagonal patterns embedded in $A^-$. This family is further divided into circular (\S\ref{sec:circhex}) and planar (\S\ref{sec:planhex}) localized hexagonal patterns. We distinguish between two cases: the supercritical case $\gamma<\gamma^T$ and the subcritical case $\gamma^T<\gamma<\gamma^F_H$. As in \S\ref{sec:locring} we fix $\nu=7$ and vary $\gamma$; the Turing bifurcation point is $\gamma^T=3.0845$.

\subsection{Circular localized hexagonal patterns}\label{sec:circhex}

\begin{figure}
\center
\subfigure[] {\label{fig:HEX_circhex_g2.8989_t300} \mbox{\includegraphics[width=0.4\textwidth]{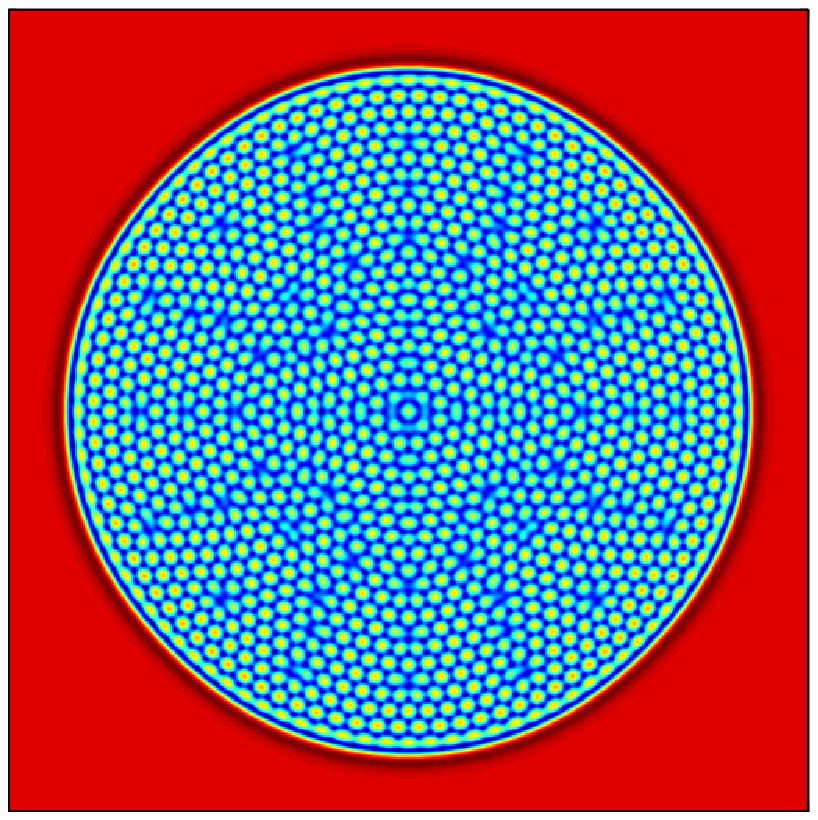}}}
\subfigure[] {\label{fig:HEX_circhex_g2.8989_t450} \mbox{\includegraphics[width=0.4\textwidth]{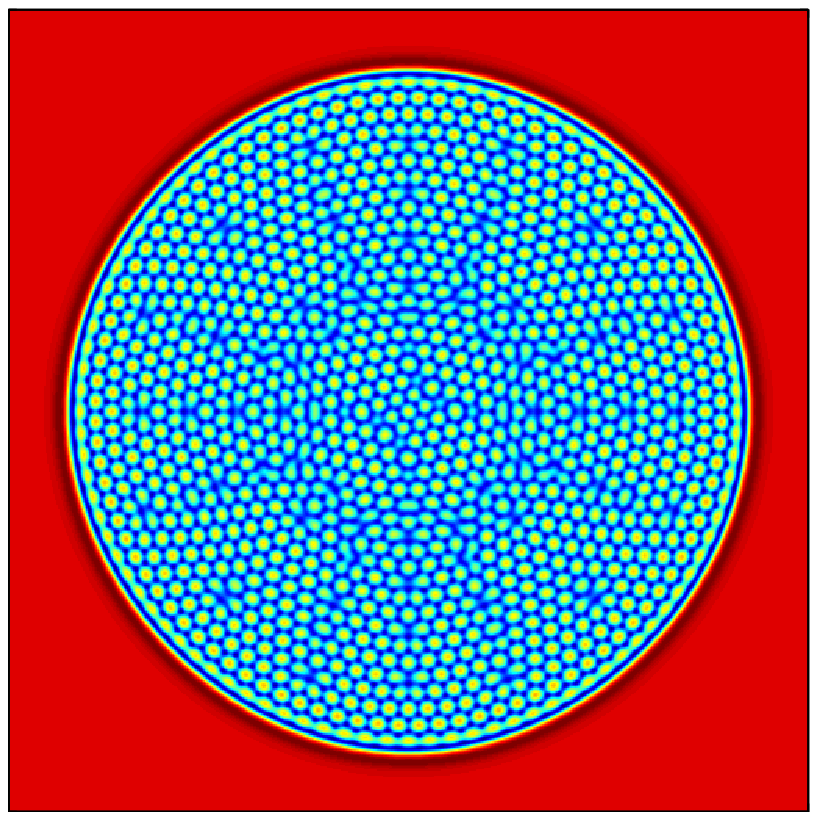}}}
\caption{Snapshots of $V(x,y)$ at (a) $t=300$ and (b) $t=450$, showing the initial evolution of circular localized hexagons. Parameters: $\nu=7$, $\gamma=2.8989$.} \label{fig:HEX_circhex_initial}
\end{figure}

In the supercritical case fully 2D LS can be found by time-evolving localized target patterns in 2D. The rings rapidly break up into hexagons as a result of azimuthal instabilities triggered by the discretization. Figure~\ref{fig:HEX_circhex_initial} shows snapshots of the initial stages of this 2D time evolution at $\gamma=2.8989$ with the localized target pattern in Fig.~\ref{fig:HEX_circhex_g2.8989_t0} used as initial condition. The fully 2D pattern that emerges consists of an approximately hexagonal pattern filling the interior of a circular front. Since the 2D solution is discretized along the $x$ and $y$ directions, the initial pattern contains multiple defects arranged into a $\mathbb{D}_4$ symmetric configuration (Fig.~\ref{fig:HEX_circhex_g2.8989_t300}). Shortly thereafter, this $\mathbb{D}_4$ symmetric pattern breaks up from the center and the defects reorganize into a pair of parabolas with $\mathbb{D}_2$ symmetry (Fig.~\ref{fig:HEX_circhex_g2.8989_t450}). This symmetry breaking process is reminiscent of the creation of $\mathbb{D}_2$ symmetric dimers or $\mathbb{D}_3$ symmetric trimers from a $\mathbb{D}_6$ symmetric hexagonal lattice of oscillons observed in experiments~\cite{UMS:96,LHARF:99}.

The subsequent evolution of the state in Fig.~\ref{fig:HEX_circhex_initial}(a) is driven by the presence of the defects. These take the form of the penta-hepta defects familiar from other haxagon-forming systems (these are located along the $x$-$y$ axes in the figure and nearer the center); however, free pentagonal defects are found near the periphery of the structure where curvature effects are significant. The latter serve as loci of phase slips whereby new cells are inserted (or deleted) from the structure leading to readjustment of nearby cells that in turn triggers the formation of new defects. As a result the whole structure undergoes slow, phase-slip driven evolution. Figure~\ref{fig:HEX_circhex_initial}(b) shows a later state of the system: the structure has almost the same size and shape but the defect distribution differs, with the defects forming prominent chains at later times (see the video at \href{\HEXcirchexa}{this URL}). In contrast to \S\ref{sec:circfr-tcf} and \S\ref{sec:locring-depin}, this defect-driven evolution cannot be axisymmetric, and so the front bounding the hexagons does not remain perfectly circular at all times. Simulations indicate that there is an approximate critical radius $\rho_0$ such that the bounding front contracts (expands) when the front radius $\rho<\rho_0$ ($\rho>\rho_0$). When $\rho$ is far from $\rho_0$, the azimuthally-averaged front speed $c$ depends mainly on $\gamma$ and the azimuthally-averaged front radius $\rho$ evolves in a manner similar to the circular traveling fronts described in \S\ref{sec:circfr-tcf}. However, when $\rho$ is near $\rho_0$, the average front speed $c$ becomes more sensitive to the details of the hexagonal pattern, which makes it difficult to determine $\rho_0$ precisely or stabilize the evolution by feedback control.

\begin{figure}
\center
\subfigure[] {\label{fig:HEX_circhex_g2.8978_t800} \mbox{\includegraphics[width=0.4\textwidth]{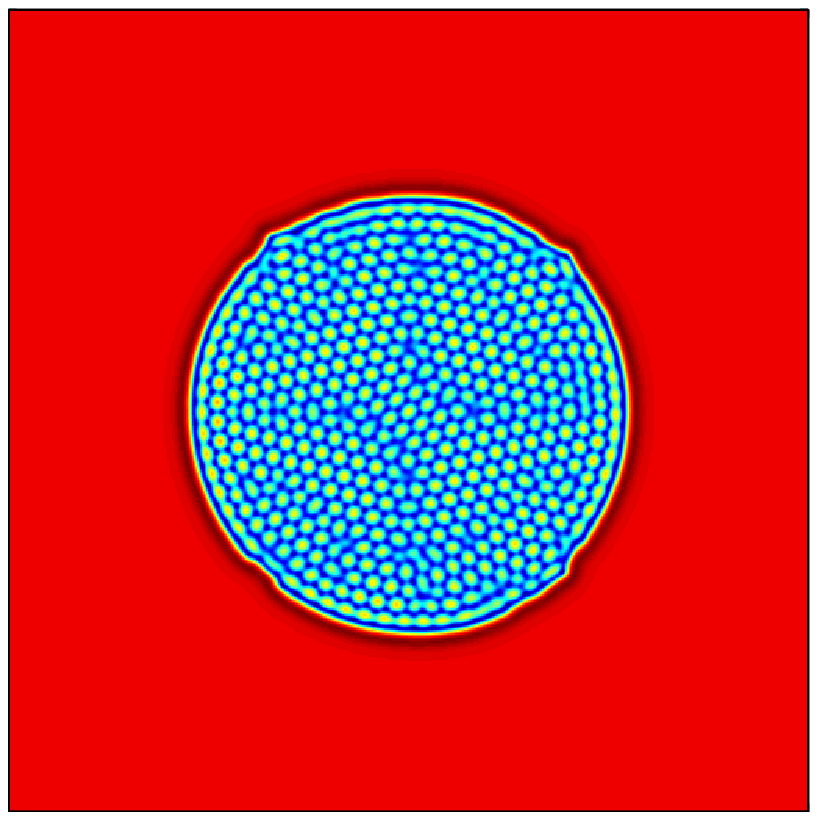}}}
\subfigure[] {\label{fig:HEX_circhex_g2.8989_t4080} \mbox{\includegraphics[width=0.4\textwidth]{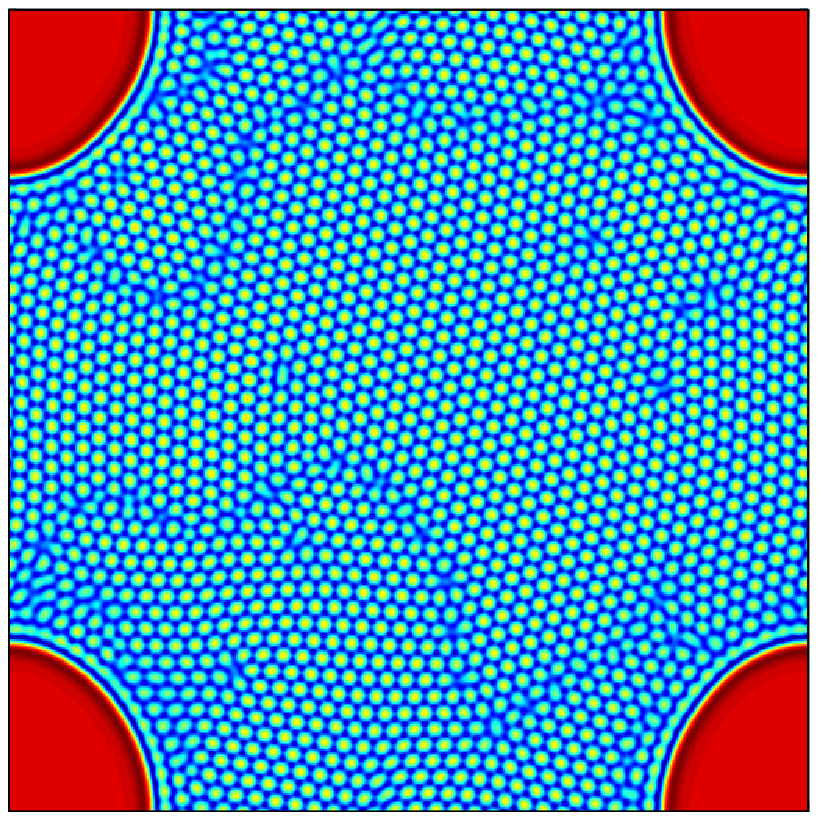}}}
\caption{Snapshots of $V(x,y)$ at $\nu=7$, and (a) $\gamma=2.8978$, $t=800$; (b) $\gamma=2.8989$, $t=4080$, showing respectively a contracting and an expanding circular localized hexagon. Videos of the evolution can be viewed at \href{\HEXcirchexs}{this URL}, and \href{\HEXcirchexa}{this URL}.} \label{fig:HEX_circhex_final}
\end{figure}

When $\rho$ is near $\rho_0$, the bounding front expands (contracts) slowly and hexagons can be created (annihilated) anywhere within the region they occupy. When $\rho$ is further from $\rho_0$, the bounding front expands (contracts) more rapidly and hexagons tend to be created (annihilated) at a fixed distance from the bounding front. This faster evolution is analogous to off-center depinning of Type-II LS in 1D. For both slow and fast evolution, the bounding front remains almost perfectly circular. However, when $\rho$ is far from $\rho_0$, fast viscous shocks may propagate azimuthally along the bounding front and these create (annihilate) hexagons on a yet faster timescale. This evolution has no 1D analog and is referred to as ultra-fast. Figure~\ref{fig:HEX_circhex_g2.8978_t800} shows a snapshot of a contracting solution obtained from an initial $\rho$ that is slightly below $\rho_0$; in this case the initial evolution is slow but transitions to fast and subsequently ultra-fast contraction once $\rho$ becomes sufficiently small. Figure~\ref{fig:HEX_circhex_g2.8989_t4080} shows a snapshot of an expanding solution for which the initial $\rho$ is slightly above $\rho_0$; in this case once fast expansion sets, the circular localized hexagonal pattern collides with its images, and evolves into a reciprocal circular localized hexagonal pattern centered on the corners. In contrast to the contraction case, ultra-fast expansion does not take place when $\rho$ becomes very large. Once the reciprocal circular localized hexagonal pattern forms, hexagons are always created right behind the bounding front; of course this front has negative curvature, in contrast to the bounding front in Fig.~\ref{fig:HEX_circhex_g2.8978_t800}. The domain is ultimately filled with a hexagonal pattern containing multiple chains of penta-hepta defects inherited from the initial evolution.

The presence of front pinning in localized target patterns and the apparent absence of pinning in circular localized hexagonal patterns can be understood by comparing the orientations of the bounding front and the periodic pattern within. In a localized target pattern, the circular front is always pinned to a periodic pattern in the direction perpendicular to itself, i.e.,~in the radial direction. However, in a circular localized hexagonal pattern, the circular front cannot be pinned to a hexagonal pattern that is only periodic along three directions.

\subsection{Interpretation: Single and double front structure}

As illustrated in Fig.~\ref{fig:HEX_elements}(b), in 2D the Turing bifurcation on $A^+$ results generically in stripes and hexagons. The bifurcation to the resulting unstable branch of stripes $S$ is supercritical as shown in Ref.~\cite{MBK10}. The bifurcation to hexagons is transcritical, with the bifurcating hexagons $H^{\pm}$ unstable on both sides of the bifurcation. As detailed next, the up-hexagons $H^{+}$ consisting of a hexagonal array of peaks (in terms of $V=\Im{(A)}$) bifurcate subcritically, i.e., into $\gamma>\gamma^T$, while the down-hexagons $H^{-}$ consisting of a hexagonal array of dips bifurcate supercritically, i.e., into $\gamma<\gamma^T$. The former turn around in a saddle-node bifurcation at $\gamma=\gamma_H^F$ and acquire stability on the upper part of the $H^+$ branch (Fig. \ref{fig:HEX_elements}(b)). Thus in $\gamma^T<\gamma<\gamma_H^F$ the stable $H^+$ state coexists with the stable $A^+$ state, while for $\gamma<\gamma^T$ only the state $H^+$ is (locally) stable and so $H^+$ coexists with the stable $A^-$ state instead.

\begin{figure}
\center
\begin{tabular}{cc}
\includegraphics[width=0.4\textwidth]{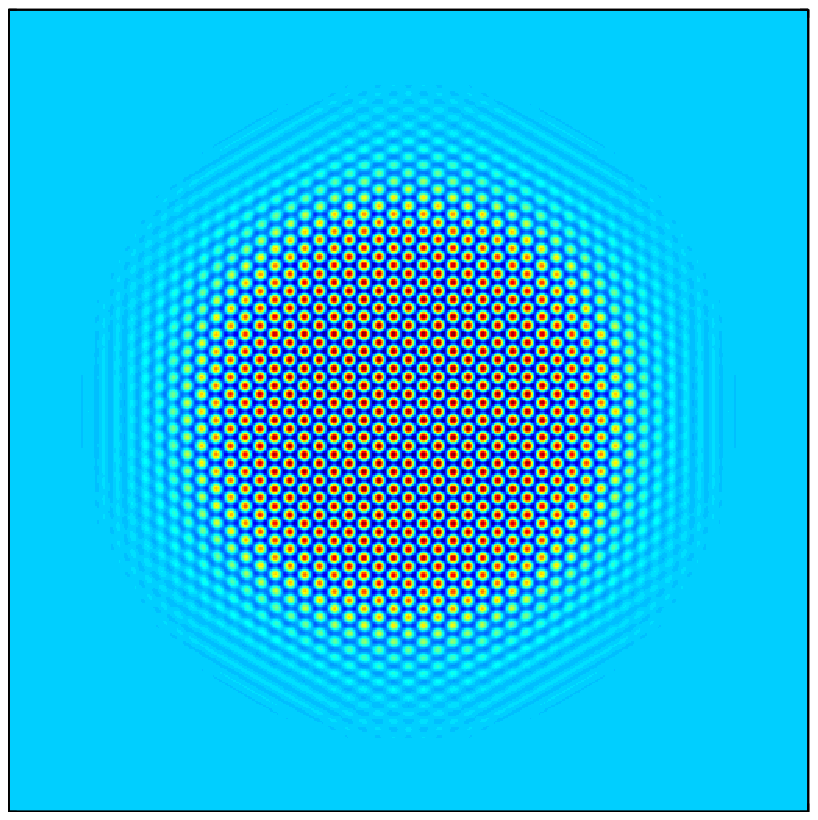} &
\includegraphics[width=0.4\textwidth]{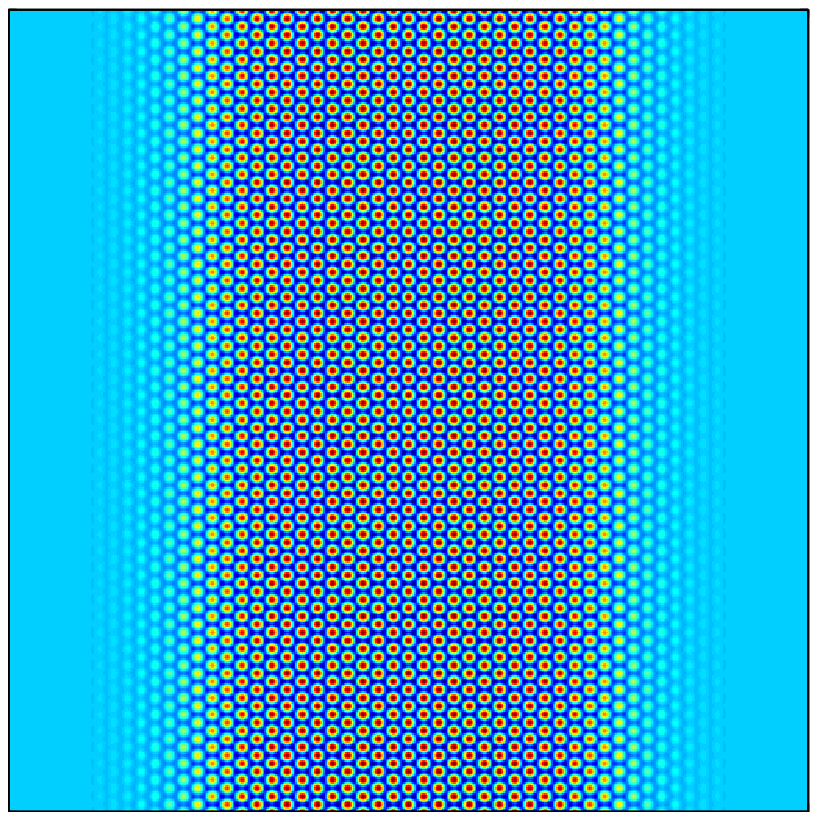} \\
(a) & (b)
\end{tabular}
\caption{Snapshots of $V(x,y)$ at (a) $\gamma=3.14405$, $t=5000$; (b) $\gamma=3.1556$, $t=10000$, showing almost steady fully and planar localized hexagonal patterns in the subcritical regime.} \label{fig:HEX_subhex}
\end{figure}

\begin{figure}
\center
\begin{tabular}{cccc}
\includegraphics[width=0.24\textwidth]{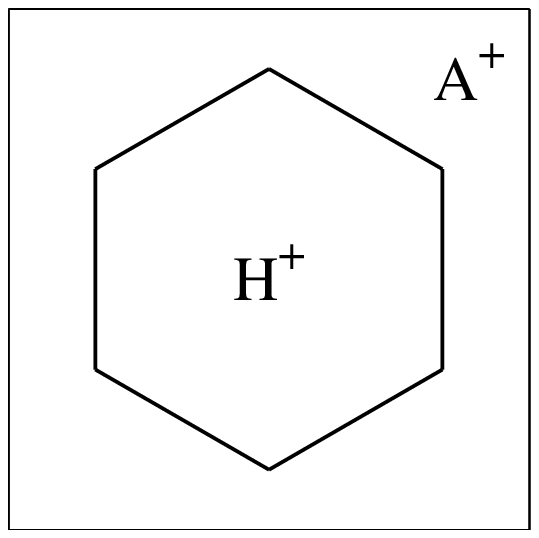} &
\includegraphics[width=0.24\textwidth]{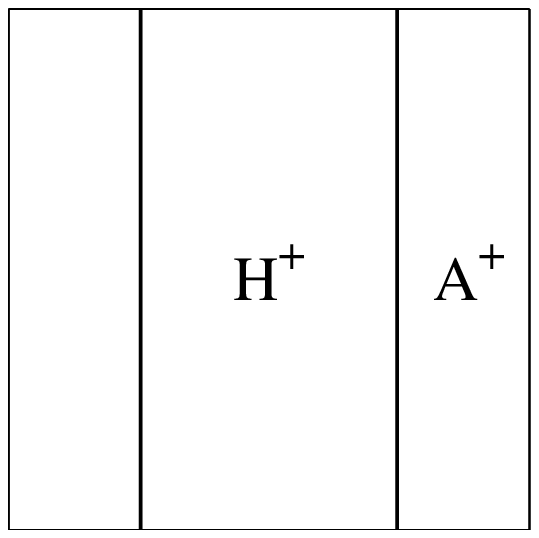} &
\includegraphics[width=0.24\textwidth]{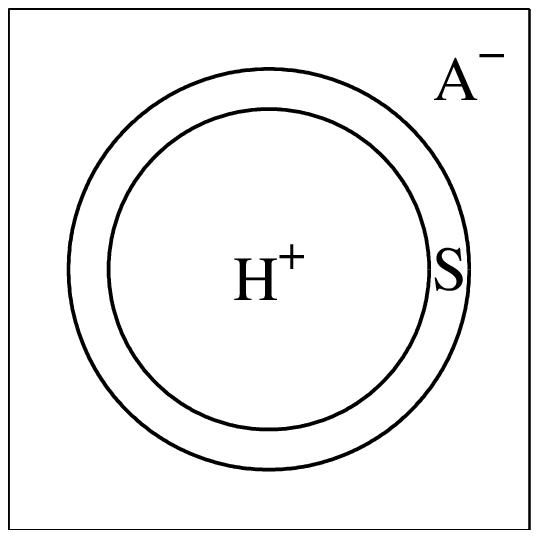} &
\includegraphics[width=0.24\textwidth]{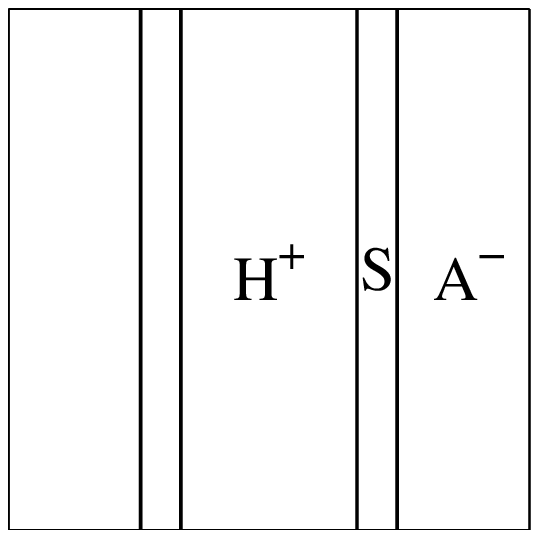} \\
(a) & (b) & (c) & (d)
\end{tabular}
\caption{Sketch of subcritical and supercritical localized hexagons, which respectively exist for $\gamma^T<\gamma<\gamma_H^F$ and $\gamma<\gamma^T$ in Fig.~\ref{fig:HEX_elements}(b). (a) Subcritical fully localized hexagons. (b) Subcritical planar localized hexagons. (c) Supercritical circular localized hexagons. (d) Supercritical planar localized hexagons.} \label{fig:HEX_lochex_illus}
\end{figure}

In the subcritical regime $\gamma^T<\gamma<\gamma_H^F$ stationary fronts exist between $H^{+}$ and $A^{+}$, and these are a consequence of the pinning of the front to the hexagonal structure on one side. As illustrated in Fig.~\ref{fig:HEX_lochex_illus}(a,b), these stationary fronts can be assembled into steady fully and planar localized hexagonal patterns.
Figure \ref{fig:HEX_subhex} shows examples of both types computed via DNS with the initial condition taken to be a hexagonal (rectangular) patch of hexagons $H^+$ embedded in $A^+$. The hexagons $H^+$ are obtained by incrementing $\gamma$ adiabatically from $\gamma<\gamma^T$ past $\gamma^T$ and into $\gamma>\gamma^T$. The computations are done with the horizontal and vertical domain sizes $L_x=200\sqrt{3}$ and $L_y=200$ for which the wavenumber $k$ is close to the critical Turing wavenumber at $\gamma^T$, and $L_y$ is occupied by exactly $N=49$ rows of hexagons. The resulting hexagons remain amplitude stable for $\gamma<\gamma_H^F$, where $\gamma_H^F\approx3.18$ and are well approximated by
\begin{equation}\label{eq:hex-cos}
A=A^++\tilde{A}(\cos{(kx)}+\cos{(k(x+\sqrt{3}y)/2)}+\cos{(k(x-\sqrt{3}y)/2)})
\end{equation}
with $A^+=1.044-0.495i$ and $\tilde{A}=-0.004+0.116i$ at $\gamma=3.14>\gamma^T$. These hexagons are thus up-hexagons $H^+$ (i.e., $\tilde{V}\equiv\Im{(\tilde{A})}>0$); indeed as seen in Figs.~\ref{fig:HEX_circhex_initial} \& \ref{fig:HEX_subhex}, $V=\Im{(A)}$ consists of a hexagonal array of peaks. We remark that for some parameters the wavenumber $k$ is found to be Eckhaus unstable. For both fully and planar localized hexagonal patterns, the front between $H^+$ and $A^+$ selects a preferred hexagonal wavelength, and the subsequent dynamics depend on the choice of $\gamma$. In the supercritical regime $\gamma<\gamma^T$ the stable $H^+$ invades the unstable $A^+$ via a pulled front. As $\gamma$ increases past $\gamma^T$, the pulled front transitions into a pushed front between the bistable $H^+$ and $A^+$. In the subcritical regime $\gamma>\gamma^T$, the pushed front expands for smaller $\gamma$ and contracts for larger $\gamma$; videos of the expansion and contraction of fully localized hexagonal patterns (Fig. \ref{fig:HEX_lochex_illus}(a)) at $\gamma=3.1438$ and $\gamma=3.1441$ can be viewed at \href{\HEXsubfullym}{this URL} and \href{\HEXsubfullyp}{this URL}, while videos of the expansion and contraction of the related planar localized hexagonal patterns (Fig. \ref{fig:HEX_lochex_illus}(b)) at $\gamma=3.1555$ and $\gamma=3.1557$ can be viewed at \href{\HEXsubplanarm}{this URL}  and \href{\HEXsubplanarp}{this URL}. Interestingly, in the later stages of expansion the hexagonal pattern has line defects along the boundaries and thus does not fill the domain perfectly. For intermediate $\gamma$ the front is expected to be pinned to the hexagonal pattern over a finite interval of $\gamma$ known as the pinning region. The above DNS results suggest that the pinning region may be very narrow, which is consistent with the slowly varying front profile and the proximity to the Turing bifurcation~\cite{KoCh13}; Fig.~\ref{fig:HEX_subhex}(a) and Fig.~\ref{fig:HEX_subhex}(b) show almost steady fully and planar localized hexagonal patterns at $\gamma=3.14405$ and $\gamma=3.1556$. Similar solutions have been studied in detail using numerical continuation for SH23~\cite{LSAC08}, while associated depinning dynamics present outside the pinning region have been studied numerically in the Lengyel-Epstein model~\cite{JPMDB94} and in a crystallization model by Archer et al. \cite{AWTK:14}.

Theory~\cite{vS03} shows that in the subcritical regime $\gamma^T<\gamma<\gamma_H^F$ the front between $H^+$ and $A^+$ is a single pushed front. However, in the supercritical regime $\gamma<\gamma^T$ the front between $H^+$ and $A^+$ generally exhibits a double front structure, with a first front connecting the stable $H^{+}$ state to the unstable $S$ state, and a second front connecting the $S$ state to the unstable $A^+$ state. In Ref.~\cite{AWTK:14} the transition between the single and double fronts near $\gamma^T$ has been studied in some detail using appropriate amplitude equations \cite{HN:00,DSSS:03}. One finds that sufficiently far from $\gamma^T$ the speed of the $S$-$A^+$ front is determined by the marginal stability criterion as appropriate for a pulled front \cite{vS03} and likewise for the $H^+$-$S$ front. However, as the distance $\gamma^T-\gamma$ is reduced, the $H^+$-$S$ speed departs from the predictions of the marginal stability theory and is determined instead nonlinearly. This is a consequence of an orbit-flip whereby the front approaches $S$ with a faster decay rate than expected generically. This transition leads to a speed that connects smoothly to the pushed front speed computed in the subcritical regime.

The circular localized hexagonal patterns described in \S\ref{sec:circhex} are supercritical localized hexagons with $H^+$ embedded in $A^-$ (see Fig.~\ref{fig:HEX_circhex_initial}). Since in this parameter regime $A^+$ is unstable and coexists with stable $A^-$, the circular fronts that form between $H^+$ and $A^-$ are strongly nonlinear and cannot be described by weakly nonlinear theory in Refs.~\cite{HN:00,DSSS:03}. We conjecture, however, following \cite{AWTK:14}, that in this case the front between $H^+$ and $A^-$ exhibits an analogous double front structure, with a first front connecting the stable $H^{+}$ state to the unstable state $S$, and a second front connecting the $S$ state to the stable $A^-$ state; see Fig.~\ref{fig:HEX_lochex_illus}(c) for an illustration. The stripe state $S$ is visible in the transition region between these two fronts as a ring enclosing the hexagonal state. The $S$-$A^-$ front is described by Eq.~(\ref{eq:FCGL_radial_only}) and studied in \S\ref{sec:locring-depin}; its speed may be inward or outward depending on $\gamma$ and the front radius $\rho$. Meanwhile, the $H^+$-$S$ speed is always outward since the stable $H^+$ state necessarily invades the unstable $S$ state, and always larger than the $S$-$A^-$ speed. As a result, a single stationary stripe is observed along the boundary of the circular localized hexagonal patterns and its speed is expected to be slaved to the slower contraction or expansion speed of the hexagonal state driven by phase slips within.

The above notions become clearer in the case of localized hexagonal patterns organized in a stripe-like structure, referred to as {\it planar localized hexagonal patterns}, as illustrated in Fig.~\ref{fig:HEX_lochex_illus}(d) and discussed next.

\subsection{Planar localized hexagonal patterns}\label{sec:planhex}

\begin{figure}
\center
\subfigure[] {\label{fig:HEX_planhex_g2.8972_t0} \mbox{\includegraphics[width=0.4\textwidth]{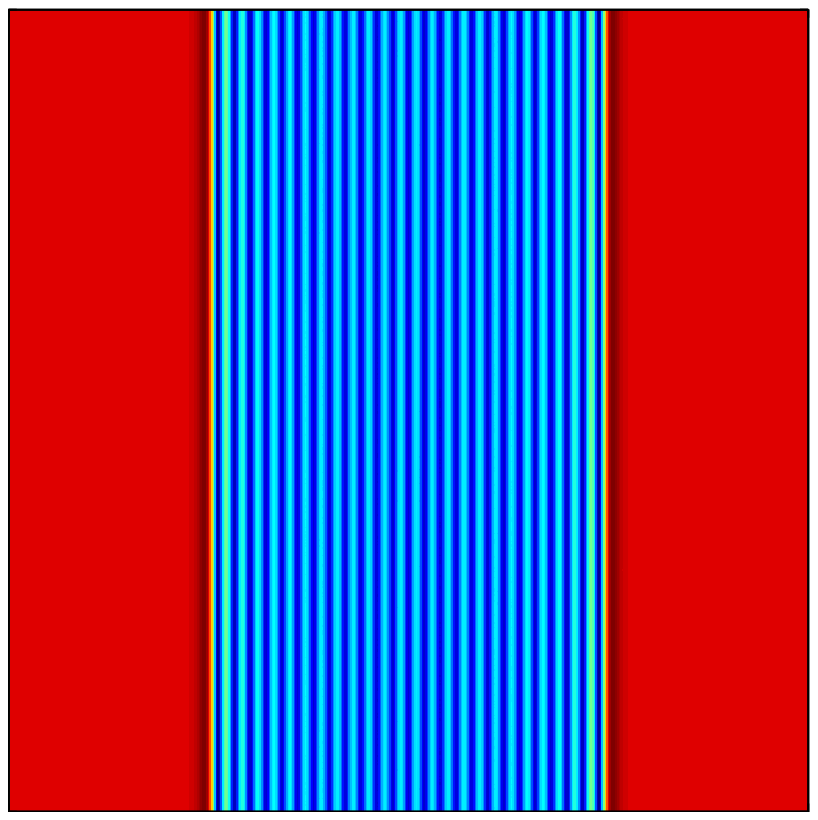}}}
\subfigure[] {\label{fig:HEX_planhex_g2.8972_t80} \mbox{\includegraphics[width=0.4\textwidth]{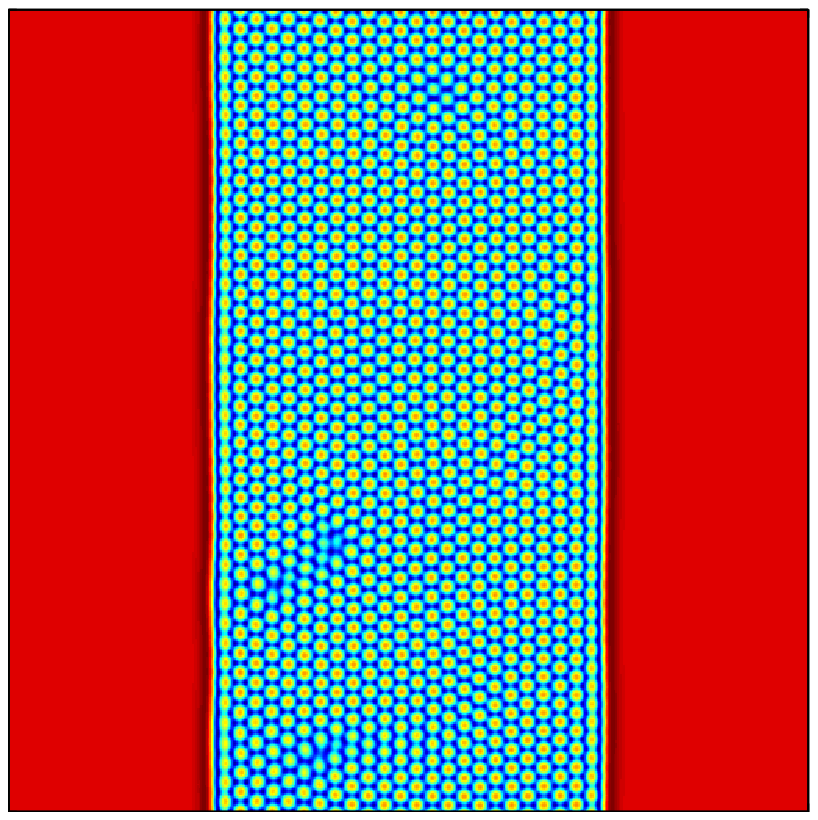}}}
\caption{Snapshot of $V(x,y)$ at (b) $t=80$, starting from localized stripes at (a) $t=0$. Parameters: $\nu=7$, $\gamma=2.8972$.} \label{fig:HEX_planhex_g2.8972_initial}
\end{figure}

Planar 2D LS can be found by time-evolving localized stripe patterns (Fig.~\ref{fig:HEX_planhex_g2.8972_t0}) in 2D, with superposed small-amplitude noise to excite transverse instabilities of the stripes. Figure~\ref{fig:HEX_planhex_g2.8972_t80} shows snapshots of the initial stages of this 2D time evolution. The fully 2D pattern that emerges consists of hexagons that align themselves along the stripes with possibly multiple defects. The subsequent dynamics of this hexagonal pattern in the $A^-$ background depend on the value of $\gamma$ used in the time evolution. However, in each case the fronts connecting the structure to $A^-$ remain approximately straight and hence the double-front picture sketched above is no longer complicated by curvature effects.

\begin{figure}
\center
\subfigure[] {\label{fig:HEX_planhex_g2.8955_t400} \mbox{\includegraphics[width=0.4\textwidth]{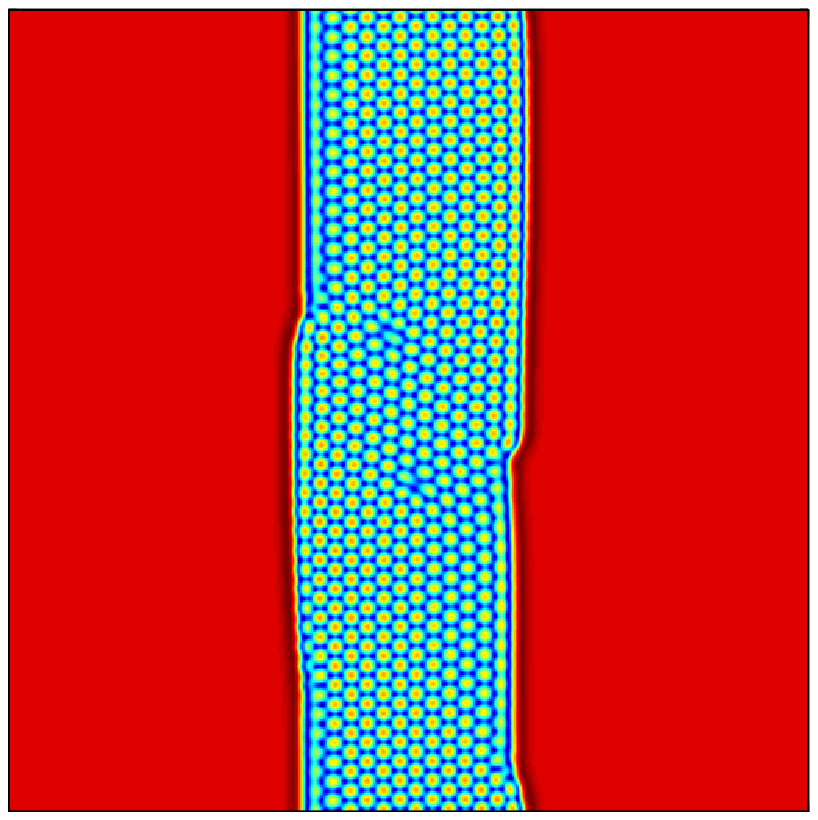}}}
\subfigure[] {\label{fig:HEX_planhex_g2.899_t1500} \mbox{\includegraphics[width=0.4\textwidth]{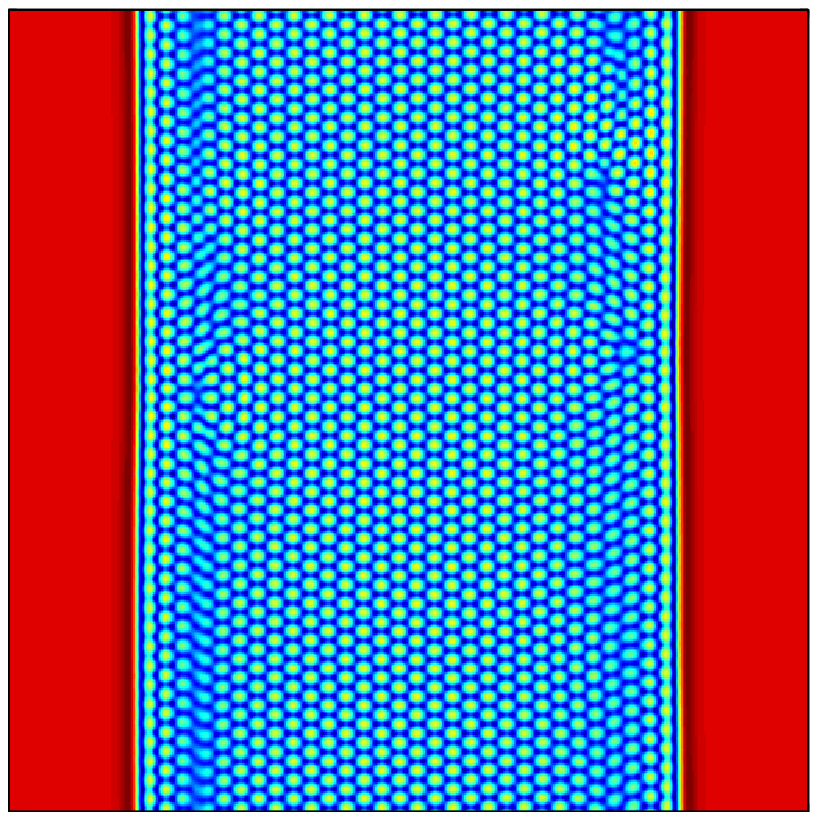}}}
\caption{Snapshots of $V(x,y)$ at $\nu=7$, and (a) $\gamma=2.8955$, $t=400$, (b) $\gamma=2.899$, $t=1500$ showing, respectively, the contraction and expansion of planar localized hexagons. Videos of the evolution can be viewed at \href{\HEXplanhexa}{this URL}  and \href{\HEXplanhexb}{this URL}. The shock-like structures propagating rapidly along the fronts are easily identified.} \label{fig:HEX_planhex_final}
\end{figure}

For smaller $\gamma$, the planar localized hexagonal pattern eventually contracts into a planar symmetric pulse, namely a 1D symmetric pulse invariant in the $y$ direction. In this process the hexagonal cells are repeatedly annihilated by viscous shocks traveling along the bounding fronts, similar to the ultra-fast contraction phase of circular localized hexagonal patterns. A snapshot of the resulting state is shown in Fig.~\ref{fig:HEX_planhex_g2.8955_t400} at $\gamma=2.8955$. For larger $\gamma$, the planar localized hexagonal pattern expands into a possibly imperfect hexagonal pattern that fills the 2D domain. In this process new hexagonal cells are created in the interior by phase slips centered on two lines of defects that result from the slow separation of the bounding fronts, similar to the fast expansion phase of circular localized hexagonal patterns. A snapshot of the resulting state is shown in Fig.~\ref{fig:HEX_planhex_g2.899_t1500} at $\gamma=2.899$.

\begin{figure}
\center
\subfigure[] {\label{fig:HEX_planhex_g2.8972_t10000} \mbox{\includegraphics[width=0.4\textwidth]{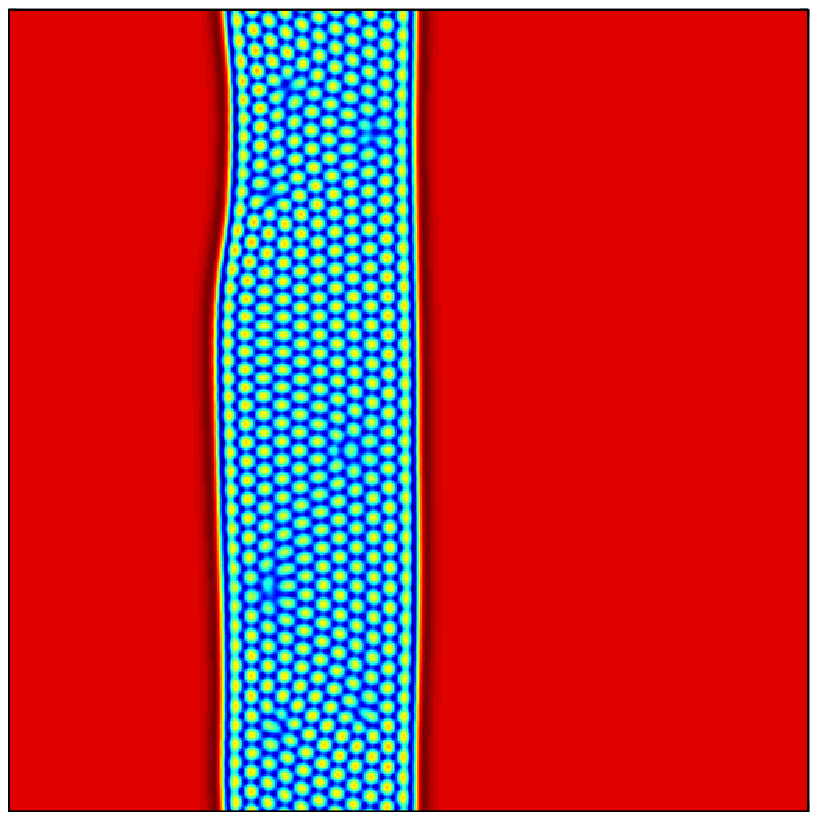}}}
\subfigure[] {\label{fig:HEX_planhex_g2.8972_t30000} \mbox{\includegraphics[width=0.4\textwidth]{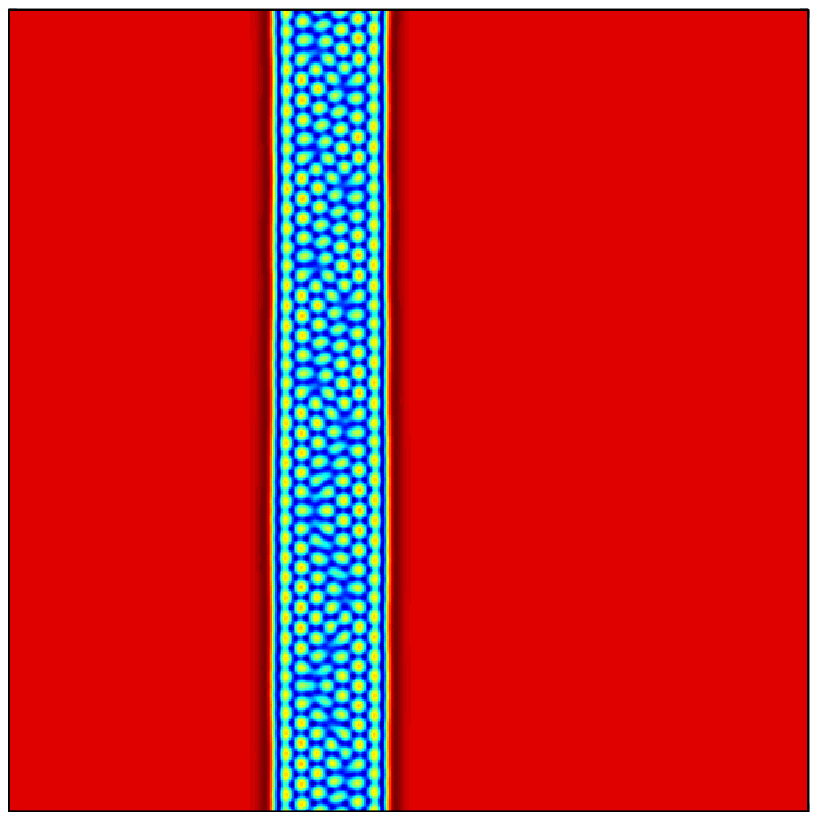}}}
\caption{Snapshots of $V(x,y)$ at (a) $t=10000$ and (b) $t=30000$ showing the competition between contraction and expansion of planar localized hexagons. Parameters: $\nu=7$, $\gamma=2.8972$. A video can be viewed at \href{\HEXplanhexc}{this URL}.} \label{fig:HEX_planhex_g2.8972}
\end{figure}

Despite the outward (inward) motion of the bounding fronts at larger (smaller) $\gamma$, we have not found steady planar localized hexagonal patterns at any intermediate $\gamma$. As shown in Fig.~\ref{fig:HEX_planhex_g2.8972_t10000}, at $\gamma=2.8972$ the boundary shocks and interior defects compete with each other, leading to non-unidirectional front motion. As a result the final state (Fig.~\ref{fig:HEX_planhex_g2.8972_t30000}) is considerably displaced from the initial centerline ($x=0$) owing to the stochastic creation and annihilation of hexagonal cells. More remarkably, this final state provides an example of a quasi-steady localized state with a complicated inner structure, which is quite unexpected in spatially homogeneous pattern forming PDEs. The inner structure consists of a few rows of hexagonal cells interspersed with defects arranged in a frustrated configuration that overall neither contracts nor expands, although the competition between these two tendencies manifests itself occasionally as local pulsations. A remarkable video (see \href{\HEXplanhexc}{this URL}) shows the evolution of this state. The earlier parts of the video show clearly the propagation of phase slips, whereby a phase slip triggers an adjacent phase slip etc. The later parts of this very long simulation show persistent dynamics in an apparently equilibrium state, i.e., a state of a more-or-less fixed width. This residual time dependence may be associated with the fact that the $H^+$ state is invading the unstable stripe state $S$ from the inside while the stable background state $A^-$ invades from the outside, just as in the case of the almost steady circular localized hexagonal patterns. The resulting fronts are approximately stationary but select wavelengths dynamically; these wavelengths may lie outside the stability balloon of the hexagonal state leading to phase slips and may, moreover, be incommensurate with the width of the structure, thereby providing an additional source of disequilibrium. Analogous planar localized hexagonal patterns in the subcritical case are shown in Fig. \ref{fig:HEX_subhex}(b).

The difference between the dynamics of circular and planar localized hexagonal patterns may be attributed to the effective boundary conditions experienced by the hexagons. In the circular case, the bounding front is a single closed curve that acts as a free boundary for the hexagons. As a result, the wavelength of the hexagons can change freely in response to front motion. However, in the planar case, the two bounding fronts are localized in $x$ and periodic in $y$, and thus act as free boundaries for the hexagons only in the $x$-direction. Thus the $x$-wavelength of the hexagons can change freely in response to front motion, but the $y$-wavelength is fixed by the periodic boundary conditions in $y$. This anisotropy of the hexagonal pattern may help explain the coexistence of contraction associated with the fronts and expansion from the interior of the structure.

\section{Conclusion}\label{sec:end-HEX}

In this paper we have characterized 2D LS in the 1:1 FCGLE with different inner patterns and front shapes. The parameters are chosen near a codimension-two point between a steady front between two spatially homogeneous equilibria and a supercritical Turing bifurcation on one of them. Steady circular fronts in the Turing-stable regime and localized target patterns in the Turing-unstable regime respectively exhibit new bifurcation structures termed radially collapsed snaking (RCS) and core-mediated snaking (CMS). Axisymmetric oscillons on the RCS and CMS branches are found to be spectrally stable over a wide parameter interval, and subject to various types of radial and azimuthal instabilities otherwise. New temporal dynamics involving front contraction or expansion are found using radial and fully 2D time evolution, the latter leading to fully 2D LS including supercritical circular and planar localized hexagonal patterns.

The stable axisymmetric oscillons in the 1:1 FCGLE may be compared with the 2:1 FCGLE~\cite{McQS14}. In the 2:1 case, stable 2D oscillons exist on a much narrower parameter interval than stable 1D oscillons. In contrast, in the 1:1 case, stable 2D oscillons exist on multiple segments of the RCS and CMS branches, and the width of the existence interval is comparable to stable 1D oscillons. The existence of multiple folds on these branches results from the existence of a supercritical Turing bifurcation near a stable front, which is not known to occur in the 2:1 case. However, stable 2D oscillons may be similarly present in other harmonically forced systems relevant to experiments.

The subcritical localized hexagonal patterns in the 1:1 FCGLE (Fig. \ref{fig:HEX_subhex}) may be compared with SH23~\cite{LSAC08}. In particular, the bifurcation diagrams of steady solutions may be computable using numerical continuation following Ref.~\cite{LSAC08}. In the supercritical case new types of temporal dynamics include the slow expansion (contraction) of the hexagonal pattern driven by phase slips, and the non-unidirectional front motion of planar localized hexagonal patterns near the equivalent of a Maxwell point. It is possible that in large domains the occurrence of phase slips may be treated as a stochastic process, with a corresponding stochastic description of the front motion. In particular, it may be that on particular timescales, the front motion near the Maxwell point can be viewed as a random walk. These features may be further understood by locating the penta-hepta defects through computing the coordination numbers of the individual hexagons, and studying their lifetimes. These ideas are left for future work.

We remark, finally, that the present system also exhibits parameter regimes where the homogeneous equilibria $A^\pm$ and the hexagonal pattern $H^+$ are all simultaneously stable. In particular, the ``Maxwell'' point between $A^\pm$ may lie inside the pinning region between $H^+$ and $A^+$. Near this tristable situation, the front speed competition may produce additional interesting dynamics.

\section*{Acknowledgements}
This research was partially supported by NSF under grant DMS-1211953 (EK).

\appendix

\section{Axisymmetric fronts in variational PDEs}\label{app:circfr-var}
In this Appendix we consider the following class of variational PDEs for $u\in\mathbb{R}$:
\begin{equation}\label{eq:var-pde}
u_t=\cL[\nabla^2]u+\cN(u),\quad\cL[\nabla^2]=\sum_i\cL_i\nabla^{2i},\quad\cN(u)=\sum_i\cN_iu^i,
\end{equation}
and seek $d$-dimensional axisymmetric traveling wave solutions of the form $u(r-ct)$, where $r\geq0$ denotes the radial coordinate. At $t=0$, these solutions satisfy the following ODE
\begin{equation}\label{eq:var-ode}
-cu_r=\cL\left[\partial_{rr}+\frac{d-1}{r}\partial_r\right]u+\cN(u).
\end{equation}
To derive an expression for the instantaneous speed $c$, we expand the operator $\cL$ as
\begin{equation}\label{eq:var-cL}
\cL\left[\partial_{rr}+\frac{d-1}{r}\partial_r\right]
=\cL[\partial_{rr}]+\frac{d-1}{r}\partial_r\cL'[\partial_{rr}]+O\left(\frac{d-1}{r^2}\right),
\end{equation}
multiply Eq.~(\ref{eq:var-ode}) by $u_r$ and integrate over $r$ to obtain
\begin{equation}\label{eq:var-ode-int}
c\int_0^\infty u_r^2dr
=-\int_0^\infty u_r\left(\cL\left[\partial_{rr}\right]u+\cN(u)\right)dr
-\int_0^\infty u_r\frac{d-1}{r}\partial_r\cL'[\partial_{rr}]udr+O\left(\frac{d-1}{r^2}\right).
\end{equation}
For the integral of $u_r^2$ to be bounded, we require $u_r(r\rightarrow\infty)=0$ or equivalently $u(r\rightarrow\infty)=u^+$, where $u^+$ is a constant.

The first term on the right side of Eq.~(\ref{eq:var-ode-int}) becomes
\begin{equation}\label{eq:var-cM}
\mathcal{M}\equiv H(r=0)-H(r=\infty),
\end{equation}
where the function $H(r)$ denotes the following indefinite integral
\begin{equation}
H(r)\equiv\int u_r\left(\cL\left[\partial_{rr}\right]u+\cN(u)\right)dr.
\end{equation}
The integrand takes the form of a total derivative. After integration, the function $H(r)$ depends only on the $j$-th derivatives $u^{(j)}(r)$, $j=0,1,2,\cdots$, and can be identified as the 1D steady state Hamiltonian of Eq.~(\ref{eq:var-pde}).

Now let us assume that $u_r$ nearly vanishes except in an $O(1)$ interval around $r=\rho$ for $\rho\gg1$, or equivalently that $u(r)$ is an axisymmetric front with large radius $\rho$ connecting two flat states $u(r\rightarrow0)=u^-$ and $u(r\rightarrow\infty)=u^+$. In this case we can simplify the second term on the right side in Eq.~(\ref{eq:var-ode-int}) by approximating $r^{-1}$ with $\rho^{-1}$ in the integrand and then integrating by parts to obtain
\begin{equation}\label{eq:var-cK}
\mathcal{K}\equiv\frac{d-1}{\rho}\int_0^\infty u_{rr}\cL'[\partial_{rr}]udr,
\end{equation}
with an error term $O((d-1)\rho^{-2})$ that can be combined with the third term of Eq.~(\ref{eq:var-ode-int}).

In summary, we obtain from Eq.~(\ref{eq:var-ode-int}) the following analytic expression for the front speed $c$
\begin{equation}\label{eq:var-c-circfr}
c=\left(\int_0^\infty u_r^2dr\right)^{-1}\left(\mathcal{M}+\mathcal{K}\right)+O\left(\frac{d-1}{\rho^2}\right),
\end{equation}
where $\mathcal{M}$ and $\mathcal{K}$ are given by Eqs.~(\ref{eq:var-cM}) and (\ref{eq:var-cK}). Since for large $\rho$ the $r^{-1}\partial_r$ term in Eq.~(\ref{eq:var-ode}) can be omitted without affecting the relevant dynamics to leading order, the integrals in Eqs.~(\ref{eq:var-cK}) -- (\ref{eq:var-c-circfr}) can be approximated by replacing $u(r)$ with the 1D front profile at the same parameters. Physically the front motion is driven by both the free energy (or equivalently Hamiltonian) difference $\mathcal{M}$ and the surface (or interfacial) energy $\mathcal{K}$. To leading order $\mathcal{M}$ evaluates to
\begin{equation}
\mathcal{M}_0\equiv H(u^-)-H(u^+),
\end{equation}
and may take two qualitatively different forms:
\begin{itemize}
\item The two equilibria $u^\pm$ are equivalent when $u(r)$ is a pulse ($u^-=u^+$) or an Ising front ($u^-=-u^+$ and $\cN(-u)=-\cN(u)$). In this case
\begin{equation}\label{eq:var-cM-eq}
\mathcal{M}_0=0.
\end{equation}
\item The two equilibria $u^\pm$ are inequivalent when $u(r)$ is a Bloch front. In this case $\mathcal{M}_0=0$ only at the Maxwell point $\gamma=\gamma_M$ with $\gamma$ denoting the control parameter. Near $\gamma_M$
\begin{equation}\label{eq:var-cM-neq}
\mathcal{M}_0\propto \gamma-\gamma_M.
\end{equation}
\end{itemize}
The term $\mathcal{K}$ is proportional to the front curvature $\kappa\equiv\rho^{-1}$. It follows from Eq.~(\ref{eq:var-c-circfr}) and the definition $c=d\rho/dt$ that steady axisymmetric fronts are unstable with respect to radius change when $\mathcal{K}/\kappa<0$, and stable otherwise. In particular, for one-component reaction-diffusion equations, or equivalently when $\cL[\nabla^2]=\nabla^2$, these fronts are always unstable because
\begin{equation}
\mathcal{K}/\kappa=-(d-1)\int_0^\infty u_r^2dr<0.
\end{equation}
However, when higher order terms like $\nabla^4$ are present in $\cL[\nabla^2]$, as is the case for the Swift-Hohenberg equation and its generalizations, these fronts may be stable in certain parameter regimes since $\mathcal{K}/\kappa$ is an alternating sum that can take either sign.

\subsection{Radially collapsed snaking in $(1+\epsilon)$ dimensions}
The function $c(\rho)$ determines the front evolution via $d\rho/dt=c(\rho)$; this function in turn depends on $\mathcal{M}$ and $\mathcal{K}$. The term $\mathcal{M}$ depends on $\rho$ due to the deviation of $u(r)$ from $u^-$ near $r=0$, i.e.~the local front profile. This profile is monotonic (oscillatory) in $r$ if the slowest unstable eigenvalue(s) of $u^-$ are real (complex). Hereafter we consider only the complex case, which occurs when $\cL[\nabla^2]$ contains $\nabla^4$ or higher order terms; the analysis in the real case is similar. We introduce $\tilde{u}(r)\equiv u(r)-u^-$, and denote the relevant eigenvalues as $\lambda_r\pm i\lambda_i$, where $\lambda_{r,i}>0$, assuming that $\lambda_i=O(1)$.

When $d=1$, $\tilde{u}^{(j)}(r=0)$ vanishes for $j$ odd and scales as
\begin{equation}\label{eq:rcs-1pe-uscale}
\tilde{u}^{(j)}(r=0)\sim\exp{(-\lambda_r\rho)}\sin{(\lambda_i\rho+\phi)}
\end{equation}
for $j$ even, where $\phi$ is a constant. Since $H(r=0)-H(u^-)$ is a quadratic form in $\tilde{u}^{(j)}(r=0)$, the higher order correction to $\mathcal{M}$ due to the oscillatory tail scales as
\begin{equation}\label{eq:rcs-1pe-Mscale}
\mathcal{M}-\mathcal{M}_0\sim\exp{(-2\lambda_r\rho)}\sin{(2\lambda_i\rho+\psi)},
\end{equation}
where $\psi$ is a constant. Hence $c\sim\mathcal{M}$ is a decaying oscillatory function of $\rho$; this behavior is known as collapsed snaking (CS)~\cite{KW:05}.

For $d>1$, $\mathcal{M}$ still exhibits oscillatory decay with $\rho$, but with a different scaling from Eq.~(\ref{eq:rcs-1pe-Mscale}) unless $d-1\ll1$. The contribution from $\mathcal{K}$ is monotonic in $\rho$ and scales as
\begin{equation}
\mathcal{K}\sim(d-1)\rho^{-1}.
\end{equation}
If $\lambda_r=O(1)$ and $d-1=O(1)$, then $c\sim\mathcal{M}+\mathcal{K}$ decays monotonically in $\rho$ for $\rho\gg1$. However if either $\lambda_r\ll1$ or $d-1\ll1$, then there exists a critical radius $\rho^{RCS}\gg1$ such that $c$ is monotonic in $\rho$ for $\rho>\rho^{RCS}$ and oscillatory in $\rho$ for $\rho<\rho^{RCS}$; this is radially collapsed snaking (RCS). The critical radius $\rho^{RCS}$ satisfies $d(\mathcal{M}+\mathcal{K})/d\rho=0$, i.e.
\begin{equation}\label{eq:var-RCS-eq}
\exp{(-2\lambda_r\rho)}\sim(d-1)\rho^{-2}.
\end{equation}
In \S\ref{sec:circtw} an instance of RCS is found numerically at $d=2$ and $\lambda_r\ll1$ in the 1:1 FCGLE, or specifically Eq.~(\ref{eq:circfr_ode}). It is an open question whether $\lambda_r\ll1$ is possible in Eq.~(\ref{eq:var-ode}). In the $d-1\ll1$ case, denoting $\epsilon\equiv d-1$, we can solve Eq.~(\ref{eq:var-RCS-eq}) asymptotically, obtaining
\begin{equation}\label{eq:rho-RCS-1pe}
\rho^{RCS}=\frac{1}{2\lambda_r}(-\log{\epsilon}+2\log{(-\log{\epsilon})})+O(1).
\end{equation}

\begin{figure}
\centering \includegraphics[width=0.48\textwidth]{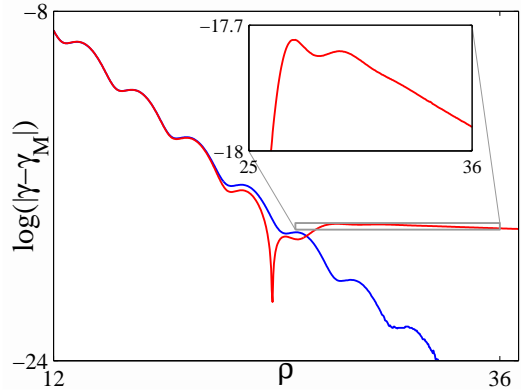}
\caption{The logarithm of the distance from the Maxwell point $|\gamma-\gamma_M|$ as a function of the front radius $\rho$ computed from numerical continuation of axisymmetric fronts in SH23, namely Eq.~(\ref{eq:sh23}). Parameters: $b_2=2.7$, $d-1=0$ (blue), $d-1=10^{-6}$ (red).} \label{fig:HEX_sh23_rcs}
\end{figure}

To take a concrete example, we consider the steady-state axisymmetric quadratic-cubic Swift-Hohenberg equation (SH23) in $d$ dimensions, namely
\begin{equation}\label{eq:sh23}
0=\left[\gamma-\left(\partial_{rr}+\frac{d-1}{r}\partial_r+1\right)^2\right]u+b_2u^2-u^3,
\end{equation}
where $b_2$ and $\gamma$ are constants. At $b_2=2.7$, the Maxwell point occurs at $\gamma_M=-0.62$ between $u^-=1.8$ and $u^+=0$, and the real part of the unstable eigenvalues of $u^-$ is $\lambda_r=0.37$. Figure~\ref{fig:HEX_sh23_rcs} shows the logarithm of the distance from the Maxwell point, $|\gamma-\gamma_M|$, as a function of the front radius $\rho$ computed from numerical continuation of axisymmetric fronts between $u^\pm$ at $d-1=0$ (blue) and $d-1=10^{-6}$ (red). In the $d-1=0$ case, the branch undergoes CS with decay exponent $-2\lambda_r$. In the $d-1=10^{-6}$ case, the branch undergoes RCS with oscillatory and monotonic dependence on $\rho$ respectively for $\rho<\rho^{RCS}$ and $\rho>\rho^{RCS}$. The predicted critical radius from Eq.~(\ref{eq:rho-RCS-1pe}) is $\rho^{RCS}=26+O(1)$, while the predicted radius at which $\gamma=\gamma_M$ is $\rho=22+O(1)$, both of which agree well with the numerical computations.

\bibliographystyle{elsarticle-num}
\biboptions{compress}

\end{document}